\documentclass[]{interact}

\usepackage{adjustbox}

\usepackage{amsmath}
\usepackage{multirow}

\usepackage[mathscr]{euscript}
\usepackage{epstopdf}
\usepackage[caption=false]{subfig}

\usepackage[numbers,sort&compress]{natbib}
\bibpunct[, ]{[}{]}{,}{n}{,}{,}
\renewcommand\bibfont{\fontsize{10}{12}\selectfont}
\makeatletter
\def\NAT@def@citea{\def\@citea{\NAT@separator}}
\makeatother

\theoremstyle{plain}

\theoremstyle{definition}

\theoremstyle{remark}

\begin{document}


\title{Application of Dynamic Linear Models to Random Allocation Clinical Trials with Covariates}

\author{
\name{Albert H. Lee III\thanks{CONTACT Albert H. Lee III. Email: leeah2@vcu.edu}}
\affil{Virginia Commonwealth University, Richmond, Virginia, USA}
}

\maketitle

\begin{abstract}
A recent method using Dynamic Linear Models to improve preferred treatment allocation budget in random allocation models was proposed by \cite{Lee2020DLMRAC}. However this model failed to include the impact covariates such as smoking, gender, etc, had on model performance. The current paper addresses random allocation to treatments using the DLM in Bayesian Adaptive Allocation Models with a single covariate. We show a reduced treatment allocation budget along with a reduced time to locate preferred treatment. Furthermore, a sensitivity analysis is performed on mean and variance parameters and a power analysis is conducted using Bayes Factor. This power analysis is used to determine the proportion of unallocated patient budgets above a specified cutoff value.  Additionally a sensitivity analysis is conducted on covariate coefficients. 
\end{abstract}

\begin{keywords}
Bayes Factor, Dynamic Linear Model, Random Allocation, Clinical Trials, Time Series
\end{keywords}

\section{Introduction}

Clinical trials are popular research methods used to determine a preferential treatment when more than one possible treatment exists by reducing between group bias. These treatments are randomly assigned to groups of patients receiving a particular treatment. According to \citet{zelen1969play} these groups \lq\lq{are as similar as possible except for the administered treatment whereby the groups are decided through randomization\rq\rq}. Randomization procedures in clinical trials have been extensively researched, and while assigning an equal number of patients to each treatment is the most common method, ethical issues using this method were discussed by \cite{ivanova2003play}.

Ideally, a sequential allocation of patients to treatments through a random method which skews patients to the most effective treatment while retaining a fully randomized process is preferred. This process, known as random allocation, has been extensively researched. This research includes the early works of  \cite{thompson1933likelihood,anscombe1963sequential,colton1963model}. Further research led to the Play the Winner Rule of \cite{robbins1952some}, and its modifications made by \cite{wei1978randomized}. Additional works include those of \cite{ivanova2003play, wei1979generalized}. A Bayesian approach was used by \cite{sabo2017optimal} to compare the works of both \cite{rosenberger2001optimal} and \cite{thall2007practical} for binary outcomes. 

Another method which has been used in random allocation processes involves adaptively allocating subjects between treatments through the Bayesian Adaptive Design. Here, Bayesian updating methods are used to allocate subjects to treatments. This design involves transforming updated information into prior information through repeated updating.  According to \citet{thall2007practical} this ability provides \lq\lq{a natural framework for making decisions based on accumulating data during a clinical trial\rq\rq}. Likewise, \citet{berry2006bayesian} indicated Bayesian updating ability provided \lq\lq{the ability to quantify what is going to happen in a trial from any point on (including from the beginning), given the currently available data\rq\rq}. There has been much research done in this area including the works of \cite{connor2013bayesian, trippa2012bayesian, zhou2008bayesian, collins2012bayesian}. Additionally, the works of \citet{sabo2014adaptive} illustrated a Bayesian approach to create what he termed  \lq\lq{Decreasingly Informative Prior\rq\rq} information. This was used to evaluate the  adaptive allocation performance when using binary variables. Recently, \cite{Lee2020DLMRAC} used a Dynamic Linear Model approach to random allocation, and demonstrated reduced time and patient budget used to identify the preferred treatment.

Often with human subjects however, there exist covariates such as smoking, age, or sex to mention a few, which may impact the response. It is therefore imperative to include these covariates, provided they exist, when randomly allocating subjects to treatments. The literature for covariate influenced adaptive allocation is quite sparse. The idea of $D_{A}$ optimality was discussed by \cite{atkinson1982optimum} for a biased coin design method, however, this did not include the random allocation. The works of \cite{zhang2006response} compared several random allocation methods, however, they did not include any covariate influences.  Although \cite{biswas2009optimal} used normal responses, they failed to consider the influence of covariates. A covariate adjusted method was proposed by \cite{zhang2009new} for the Doubly Adaptive Biased Coin Design, however, it looked at the variability reduction rather than the allocation methods. An examination of the asymptotic properties along with a theoretical examination may be seen in \cite{zhang2007asymptotic}, however, as with the previous authors, no random allocation was completed. However, \cite{bhattacharya2015class}  were able to use their method  when covariates were present. 

When investigating the impact of a single covariate, let $y$ patients enter a random allocation study sequentially at different times each with a single covariate $x$. These $y_{t}$ patients and their $x_{t}$ covariates may then be considered components of a time series. Additionally, patient budget size is set to be a total of $\mathscr{N}$ patients during the trial such that $\mathscr{T}$ 
is the index set for patient $y_{t}$ with covariate $x_{t}$ measured in a total of $\mathscr{N}$ patients. As these sequentially entering $y_{t}$ patients enter the allocation study updating procedures provide additional allocation information regarding treatment effectiveness toward the better treatment. Using a Bayesian Adaptive Design creates a Bayesian Learning Method, whereby information regarding the better treatment is learned as more patients enter the study. This information may then be applied to entering patients. For instance, increased information regarding the better treatment may be applied to patient $y_{15}$ through the updated information which occurred through patient $y_{14}$. Thus more information is known at patient $y_{15}$ than at patient $y_{14}$, and as information is updated, the Bayesian design learns the better treatment.
The aforementioned allocation method is capable of allocating subjects to treatments when these covariates exist.

\section*{Bayesian Methods}
Numerous works exist whereby Bayesian methodologies have been applied. Some of these works include theoretical texts by \cite{gelman1995bayesian} who applies Bayesian ideas to sampling methodologies. Additional works include those of  \cite{lee2012bayesian} who illustrates how to apply Bayesian methods using the R programming language in combination with a theoretical overview. A discussion on Bayesian Loss functions may be found in \cite{berger2013statistical}, while \cite{petris2009dynamic} chapter 1 provides an additional introduction.   

The basic premise surrounding Bayesian methods is known as Bayes rule, named after Rev. Thomas Bayes. 
The idea posed by Bayes was 
\begin{center}
	\begin{equation} 
	p(\theta | y) = p(\theta)p(y| \theta)/p(y)
	\end{equation} \label{eq:1}
\end{center}
where $p(\theta | y)$ represents the posterior distribution of $\theta$ given the known $y$ data.  Likewise 
\begin{center}
	\begin{equation} 
	p(\theta)p(y| \theta) \propto p(\theta,y)
	\end{equation} \label{eq:2}
\end{center}

Here, $p(\theta)$ is defined to be the prior probability of the parameter $\theta$ by \citet{gelman1995bayesian} and $p(\theta,y)$ is the conditional probability involving $\theta$ and $y$. Furthermore, by conditioning on the known $y$ data, the sampling distribution probability, $p(y| \theta)$ provides the posterior probability (See \cite{gelman1995bayesian} for more details.) Additional work using these ideas in the application of time series data has been done by \cite{petris2009dynamic}. Yet, once the posterior probability $p(\theta | y)$ has been calculated, it may then be used as a new prior probability and the process repeated, with the Bayesian Updating learning along the way.

The Dynamic Linear Model (DLM) of \citet{harrison1999bayesian} uses this updating process to create a Bayesian Learning Process. The learning ability created by this updating provides a useful mechanism whereby the DLM  may forecast the $y$ observations such that
\begin{eqnarray} 
\boldsymbol{Y_{t}} &=& \boldsymbol{F^{\rq}_{t} \theta_{t}} + \nu_{t} \\ \nonumber
\boldsymbol{\theta_{t}} &=&\boldsymbol{ G_{t} \theta_{t - t}} + \boldsymbol{\omega_{t}}\nonumber
\end{eqnarray} \label{eq:3}
where
\begin{eqnarray} 
\boldsymbol{\nu_{t} \sim N(0,V_{t})}\\ \nonumber
\boldsymbol{\omega_{t} \sim N(0,W_{t})} \nonumber
\end{eqnarray} \label{eq:4}


As defined both by \citet{harrison1999bayesian}, and previously in \cite{Lee2020DLMRAC} $\boldsymbol{\theta_{t}}$ represent the forecast parameter $\boldsymbol{F_{t}}$ 
where $\boldsymbol{F_{t}}$ is a known
$n \times r$ matrix of independent variables, $\boldsymbol{G_{t}}$ is a known $n \times n$ system matrix, $\boldsymbol{W_{t}}$ is a known $n \times n$ evolution variance matrix, and $\boldsymbol{V_{t}}$ is a known $r \times r$ observational variance matrix.
%

The prior forecast parameter $\boldsymbol{\theta_{t}}$ is found by noting $(\boldsymbol{\theta_{t-1}}|D_{t-1})\boldsymbol{\sim N(m_{t-1},C_{t-1})}$ for some mean $\boldsymbol{m_{t-1}}$ and variance matrix $\boldsymbol{C_{t-1}}$. The prior for $\boldsymbol{\theta_{t}}$  may be seen to be $(\boldsymbol{\theta_{t}}|D_{t-1}) \boldsymbol{\sim N(a_{t},R_{t})}$ whereby $\boldsymbol{a_{t} = G_{t}m_{t-1}}$ with $\boldsymbol{R_{t} = G_{t}C_{t-1}G^{\rq}_{t} + W_{t}}$. The one step ahead forecast is calculated as $(Y_{t}|D_{t-1})\sim N(f_{t},Q_{t})$. Here, $f_{t}$ is the current treatment allocation for patient $y$, while $Q_{t}$ is the forecast allocation variance for patient $y$. The posterior for $\boldsymbol{\theta_{t}}$ relies on
$(\boldsymbol{\theta_{t-1}}|D_{t-1})\boldsymbol{\sim N(m_{t},C_{t})}$
Furthermore, $\boldsymbol{m_{t} = m_{t-1} + A_{t}}e_{t}$, where $\boldsymbol{m_{t}}$ represents the current mean matrix, $\boldsymbol{C_{t} = R_{t}-A_{t}}Q_{t}\boldsymbol{A^{\rq}_{t}}$ where $\boldsymbol{C_{t}}$ is the current variance matrix, $\boldsymbol{A_{t} = R_{t}F_{t}}Q^{-1}_{t}$ where $\boldsymbol{A_{t}}$ is the adaptive coefficient, and $e_{t} = Y_{t} - f_{t}$ represents the error term.

\subsection*{Random Allocation Methods}
There have been several methods used to minimize allocation responses. One such solution was proposed by \cite{zhang2006response}, who suggested using


\begin{centering}
	\begin{eqnarray} \label{eq:5} \label{eq:5} \label{eq:5} 
	w_{A} &=&  
	\begin{cases}
	\frac{Q_{A_t}\sqrt{f_{B_t}}}{Q_{A_t}\sqrt{f_{B_t}}+Q_{B_t}\sqrt{f_{A_t}}} & \mbox{if ($f_{A_t}<f_{B_t}\mid\frac{Q_{A_t}\sqrt{f_{B_t}}}{Q_{B_t}\sqrt{f_{A_t}}}>1$)}\\
	\frac{Q_{A_t}\sqrt{f_{B_t}}}{Q_{A_t}\sqrt{f_{B_t}}+Q_{B_t}\sqrt{f_{A_t}}} &\mbox{if ($f_{A_t}>f_{B_t}\mid\frac{Q_{A_t}\sqrt{f_{B_t}}}{Q_{B_t}\sqrt{f_{A_t}}}<1$)}  \\
	\frac{1}{2} & \mbox{Otherwise}  \\
	\end{cases}   \\ 
	w_{B} &=& 1 - w_{A}  \nonumber
	\end{eqnarray} 
\end{centering}

to determine the optimally weighted allocation value solution. This solution was shown by \cite{biswas2009optimal} to be problematic because it was possible for $f_{A_t}$ or $f_{B_t}$ to be negative, therefore \citet{biswas2009optimal} proposed their optimal solution 

\begin{center}
	\begin{eqnarray} \label{eq:6} \label{eq:6} \label{eq:6}
	\omega_{A} &=& \frac{Q_{A_t}\sqrt{f_{B_t}}}{Q_{A_t}\sqrt{f_{B_t}}+Q_{B_t}\sqrt{f_{A_t}}}\\ \nonumber
	\omega_{B}&=&1-\omega_{A}\\ \nonumber
	\textnormal{where}\ \gamma_{A}&=&\Phi\left(\frac{f_{A_t}-f_{B_t}}{\sqrt{Q^2_{A_t}+Q^2_{B_t}}}\right),\gamma_{B}=\Phi\left(\frac{f_{B_t}-f_{A_t}}{\sqrt{Q^2_{A_t}+Q^2_{B_t}}}\right)
	\end{eqnarray} 
	
\end{center}
Recently, \cite{donahue2020allocation} examined how a Decreasingly Informative Prior distribution impacted the allocation using each of these equations. The DLM was applied by \cite{Lee2020DLMRAC} 
and used to compare the allocation results between the two equations using no covariate. In the current work a covariate is included and a comparison made. Because the DLM is an updating method at each value, the values for each of $f_{A_t}, f_{B_t}, Q_{A_t}, Q_{B_t}$ will change at each iteration, leading to different weight values based on the starting values. For this application, the covariate was generated as a Uniform (0,1) random variable.

\subsection*{Alogrithm}
To generate the allocation values
\begin{enumerate}
	
	\item Initiate the DLM for $\mu_{A}$, $\mu_{B}$, $\omega_{t}$, $C_{t_{A}}$, $C_{t_{B}}$, $Q_{t_A}$, $Q_{t_B}$.
	\item Identify $x_{t}$ and calculate predicted values and variances $f_{A_t}$ ($F_t = [1,0]$), $f_{B_t}$ ($F_t = [1,1, x_{t}]$), $Q_{A_t}$ and $Q_{B_t}$
	\item Compute $w_A$ and $w_B$
	\item Sample a Uniform(0,1) random variable U and compare $w_A$
	\item If $w_A < U$, allocate to Treatment A ($F_t = [1,0]$), otherwise allocate to treatment B ($F_t = [1,1,x_{t}]$)
	\item Conduct experiment and observe $y_{t}$
	\item Update the DLM and return to step 2 
\end{enumerate}

\subsection*{Simulation Study}

The seven scenarios in Table~\ref{tbl:SimulationScenarios} were investigated by \cite{donahue2020allocation} using the Decreasingly Informative Prior and then by \cite{Lee2020DLMRAC} using the DLM and including a covariate. Each group randomly allocated each scenario through 1000 simulations, and the treatment allocation probabilities, total number of allocations in each treatment group, and total number of successes was recorded. However, the current authors have only included the treatment allocation associated with the preferred treatment and these may be seen in Table~\ref{tbl:TreatmentGroupMSSwcov}. The Decreasingly Informative Prior Method of \cite{donahue2020allocation} utilized manual iterations for each iteration. This lead to  an sizable number of simulated calculation runs which lead to considerable completion times.
The method of \cite{Lee2020DLMRAC} was applied with a covariate added to the model and these times were greatly reduced. Each scenario was run using R Studio version 1.2.1335 on an ACER computer with an AMD Ryzen 5 2500U with Radeon Vega Mobile Gfx 2.00 GHz processor and 8.00 GB of RAM using Windows 10. The mean run time was approximately 120.259 seconds to completion. The lowest run time to completion was 60.960 seconds using the budget size $\mathscr{N} = 34$. The highest run time to completion was 120.690 seconds using budget size $N = 200$,
\begin{table}[hbt!]
	\begin{center} \captionsetup{justification=centering}\caption{Simulation Scenarios}
		\begin{tabular}{c c c c c c c c c | r r r r r r r r r } \hline
			
			Scenario & Differences & Standard Deviation & Planned Sample Budget\\ \hline
			1 &	0 &		20 &	128\\
			2 &	10 &		15 &	74\\
			3 &	10 &	    20 &	128\\
			4 &	10 &		25 &	200\\
			5 &	20 &	 	20 &	34\\
			6 &	20 &		25 &	52\\
			7 &	20 &		30 &	74\\

			\hline
		\end{tabular}\label{tbl:SimulationScenarios}
	\end{center}
\end{table}

\begin{figure}
	\centering
	\subfloat[Equal Allocation.]{%
		\resizebox*{5cm}{!}{\includegraphics{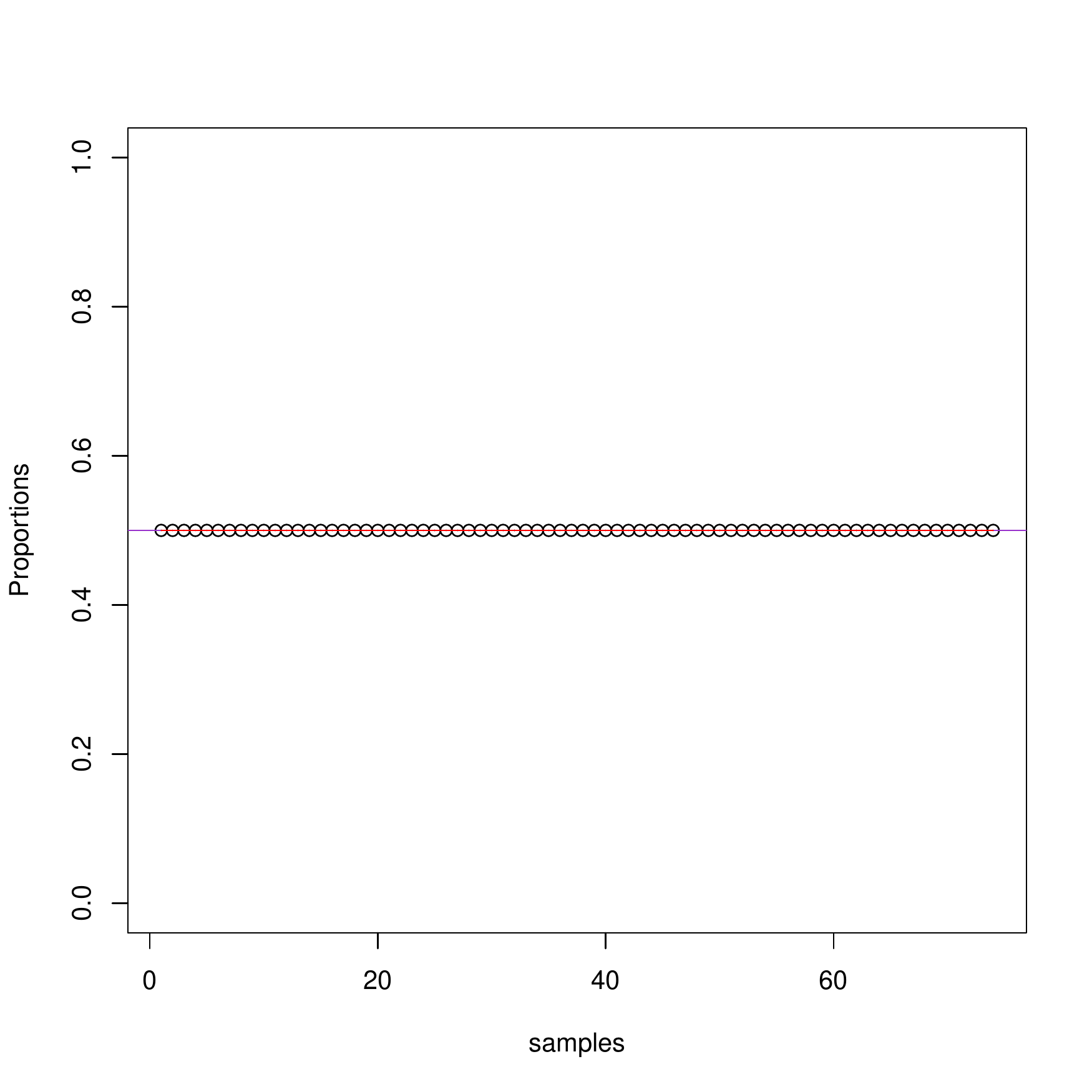}}}\hspace{5pt}
	\subfloat[Unequal Allocation.]{%
		\resizebox*{5cm}{!}{\includegraphics{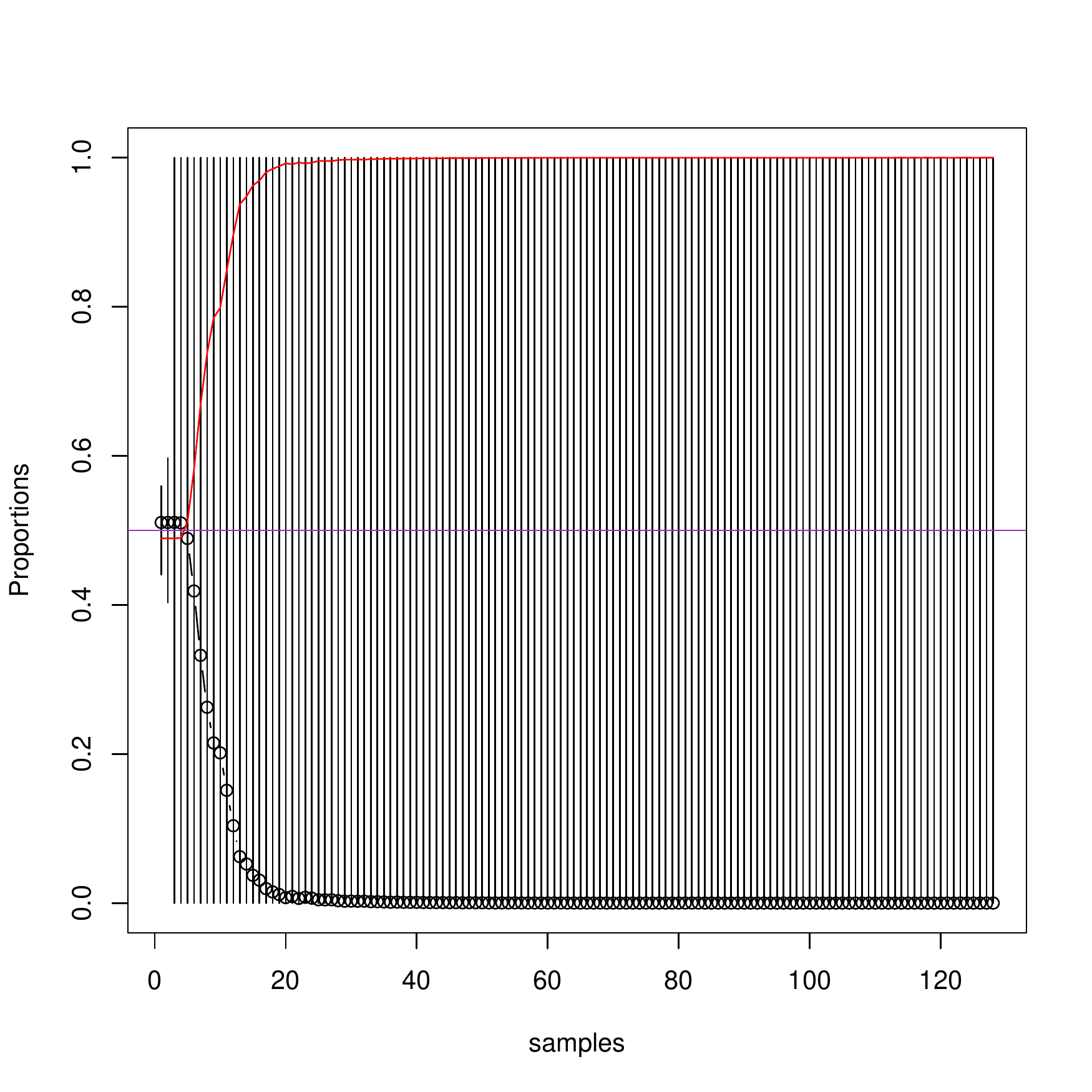}}}
	\caption{Comparison Between Equal and Unequal Allocation With Covariates.} \label{fig:AllocationFormulaWithCovariate}
\end{figure}

An analysis was conducted using each of the values in Table~\ref{tbl:SimulationScenarios} and the allocation values may be observed in Table~\ref{tbl:TreatmentGroupMSSwcov}. 
The mean number of allocations was obtained using each method. Notice the mean allocation using equation \ref{eq:5} attributed to \cite{zhang2006response} was 63.542, which is as expected, given the probability of allocation to Treatment A was 0.5. The equal treatment allocation proportion for $\mu_{B} = 0$, standard deviations  = 20 and budget size $\mathscr{N} = 128$ may be observed in Figure~\ref{fig:AllocationFormulaWithCovariate}a.  However, when the unequal method of \cite{biswas2009optimal} in equation \ref{eq:6} was applied to the same parameters, the mean number applied to Treatment A is 96.716, while the mean number allocated to Treatment B is 31.284. The proportion results for equation \ref{eq:12} may be observed in Figure~\ref{fig:AllocationFormulaWithCovariate}b. Here the mean allocation proportion for treatment A was 0.654, while mean allocation proportion to treatment B was 0.346. Additionally it is important to note the immediate convergence to either 0 or 1 using these values. Under the methods of \cite{biswas2009optimal, zhang2006response, donahue2020allocation}, the smaller value was taken to be the better allocation, therefore, it appears as though Treatment B is the favorable treatment. 

\begin{table}[ht]
	\begin{center}\caption{Treatment Group Mean Sample Size. Italicized values indicate Treatment B was selected}
		\begin{tabular}{c c c c c c } \hline
			Mean & SD&Sample&Equation \ref{eq:5}& Equation \ref{eq:6}\\
			Difference&$$&Budget&Allocation&Allocation B\\ \hline	
			0 &	20 &		128 &	$63.542$&$\textit{31.284}$\\
			10 &	15 &		74 &	$36.545$&$\textit{6.038}$\\
			10 &	20 &	    128 &	$63.690$&$\textit{8.056}$\\
			10 &	25 &		200 &	$99.064$&$\textit{10.833}$\\
			20 &	20 &	 	34 &	$16.641$&$\textit{3.314}$\\
			20 &	25 &		52 &	$25.652$&$\textit{3.900}$\\
			20 &	30 &		74 &	$36.479$&$\textit{4.379}$\\			
			
			\hline
		\end{tabular}\label{tbl:TreatmentGroupMSSwcov}
	\end{center}
\end{table}

Similar to \cite{Lee2020DLMRAC} the mean, system variance and observational variance were varied to determine treatment allocation weight behavior. The modifications made to these parameter values will aid researchers in determining a early stopping criterion through early favorable treatment identification.  This early stopping criterion will enable the avoidance of the ever-present ethical issues seen with unfavorable treatment assignment.

\begin{figure}[ht]
	\centering
	\subfloat[$\delta_t=1, \omega_t = 0.1$.]{%
		\resizebox*{4cm}{!}{\includegraphics{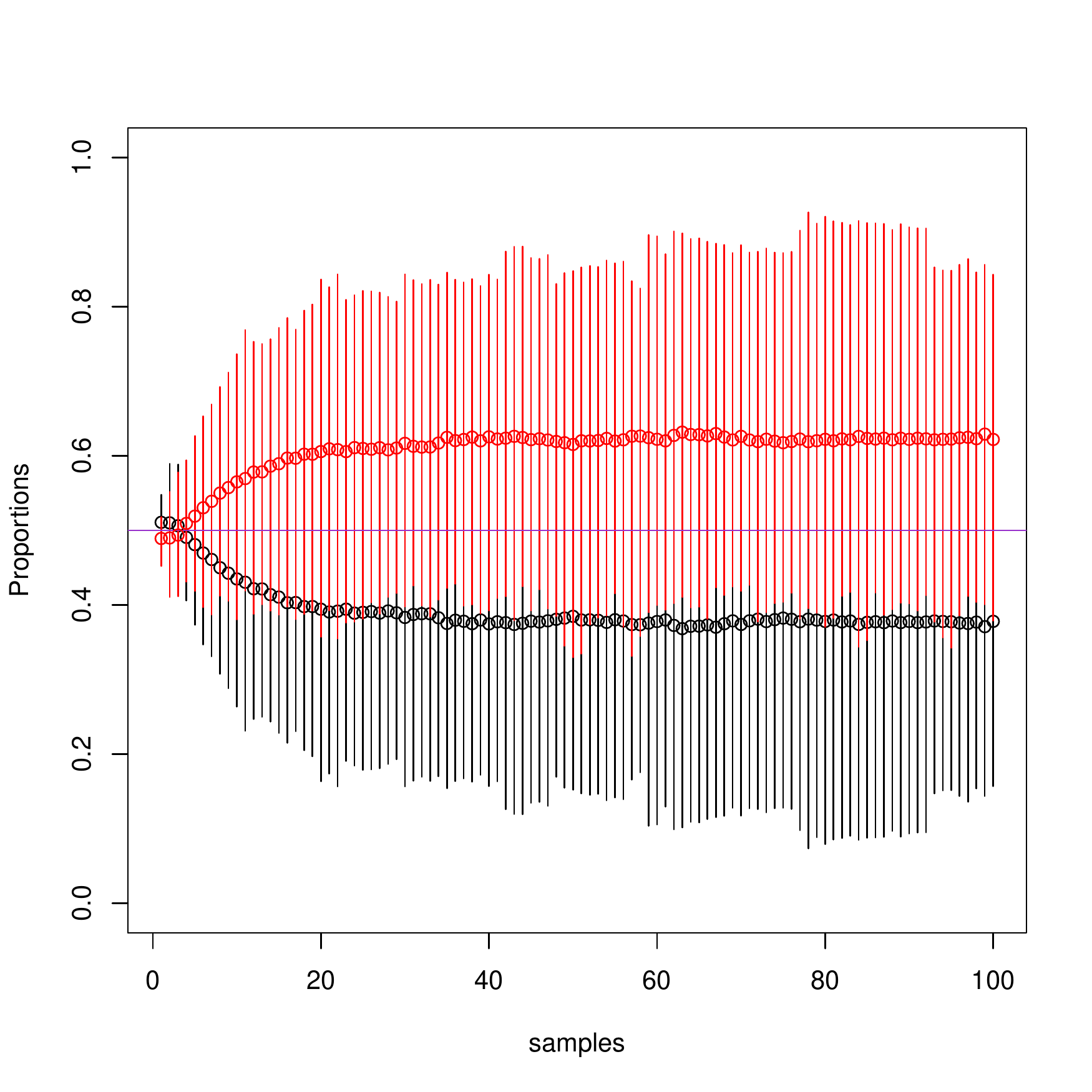}}}\hspace{5pt}
	\subfloat[$\delta_t=1, \omega_t = 0.01$.]{%
		\resizebox*{4cm}{!}{\includegraphics{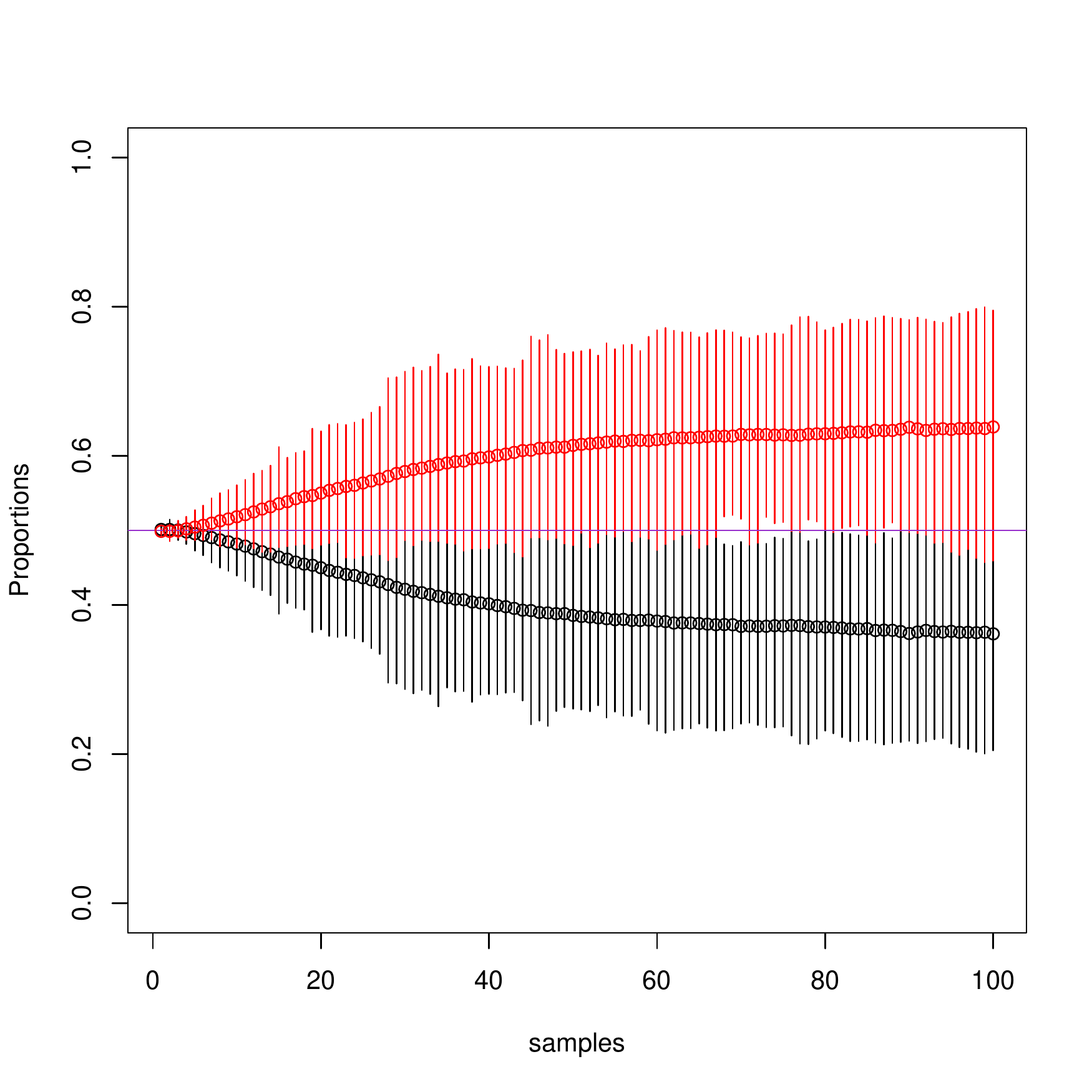}}}\hspace{5pt}
	\subfloat[$\delta_t=1, \omega_t = 0.001$.]{%
	\resizebox*{4cm}{!}{\includegraphics{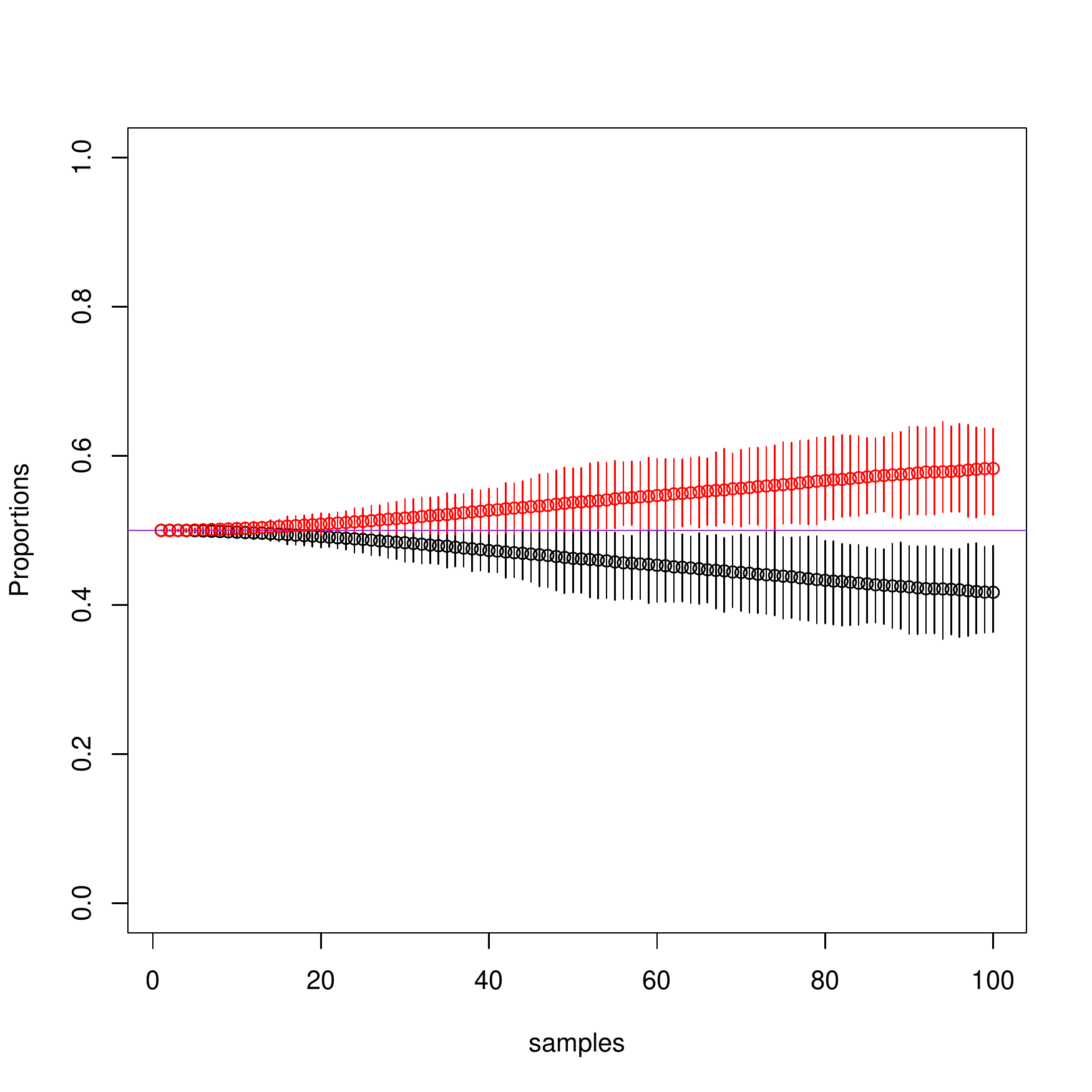}}}\\
	\subfloat[$\delta_t=3, \omega_t = 0.1$.]{%
		\resizebox*{4cm}{!}{\includegraphics{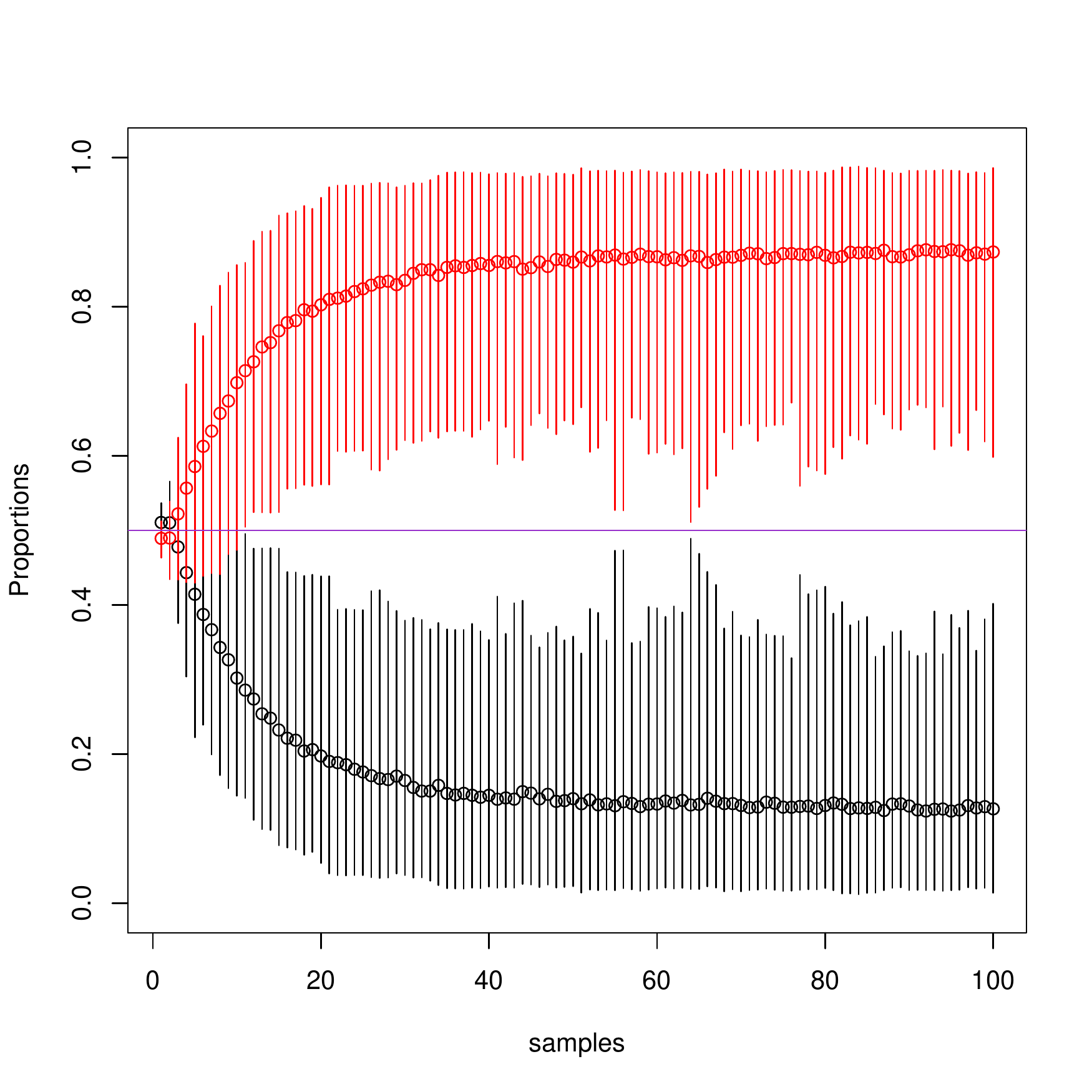}}}\hspace{5pt}
	\subfloat[$\delta_t=3, \omega_t = 0.01$.]{%
		\resizebox*{4cm}{!}{\includegraphics{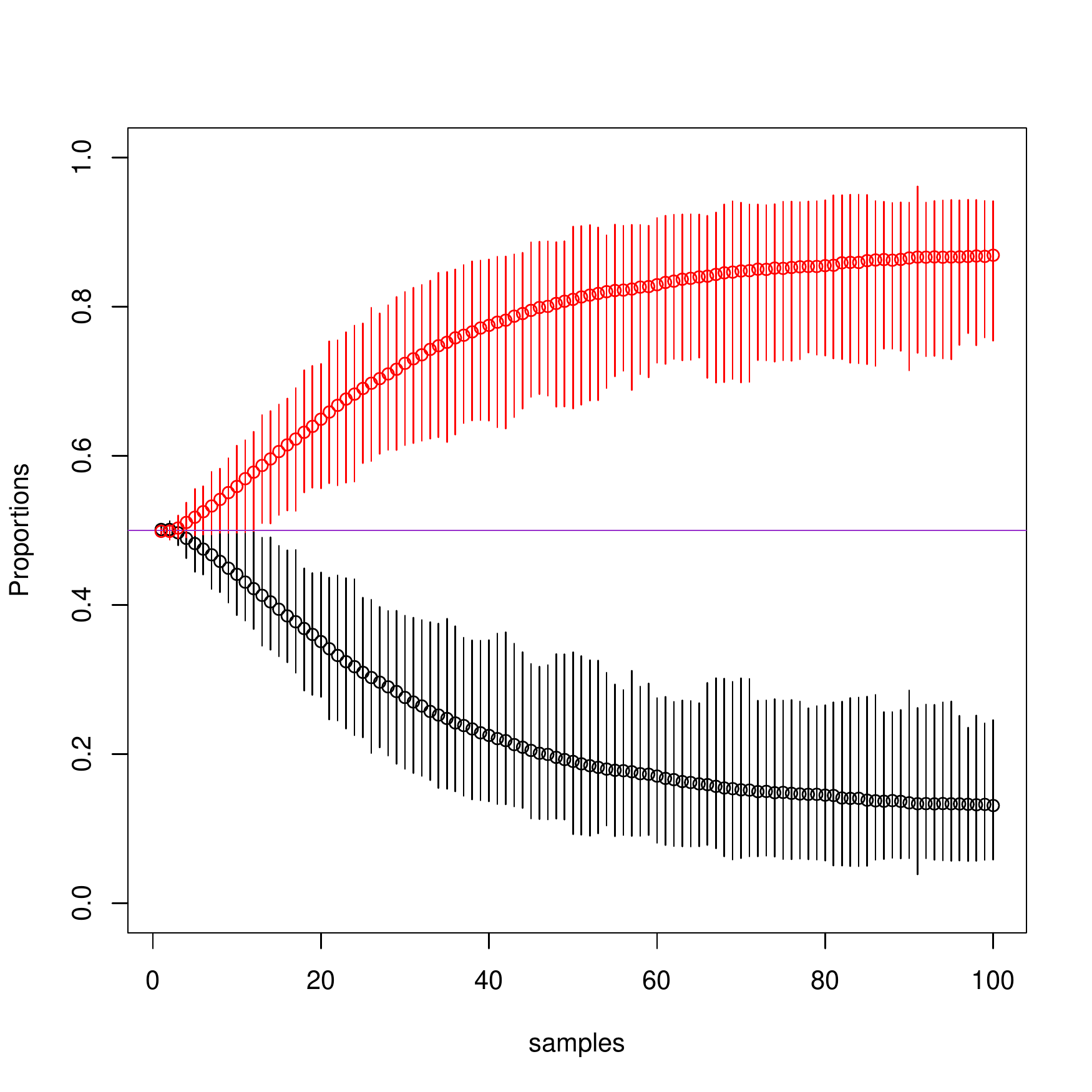}}}\hspace{5pt}
	\subfloat[$\delta_t=3, \omega_t = 0.001$.]{%
		\resizebox*{4cm}{!}{\includegraphics{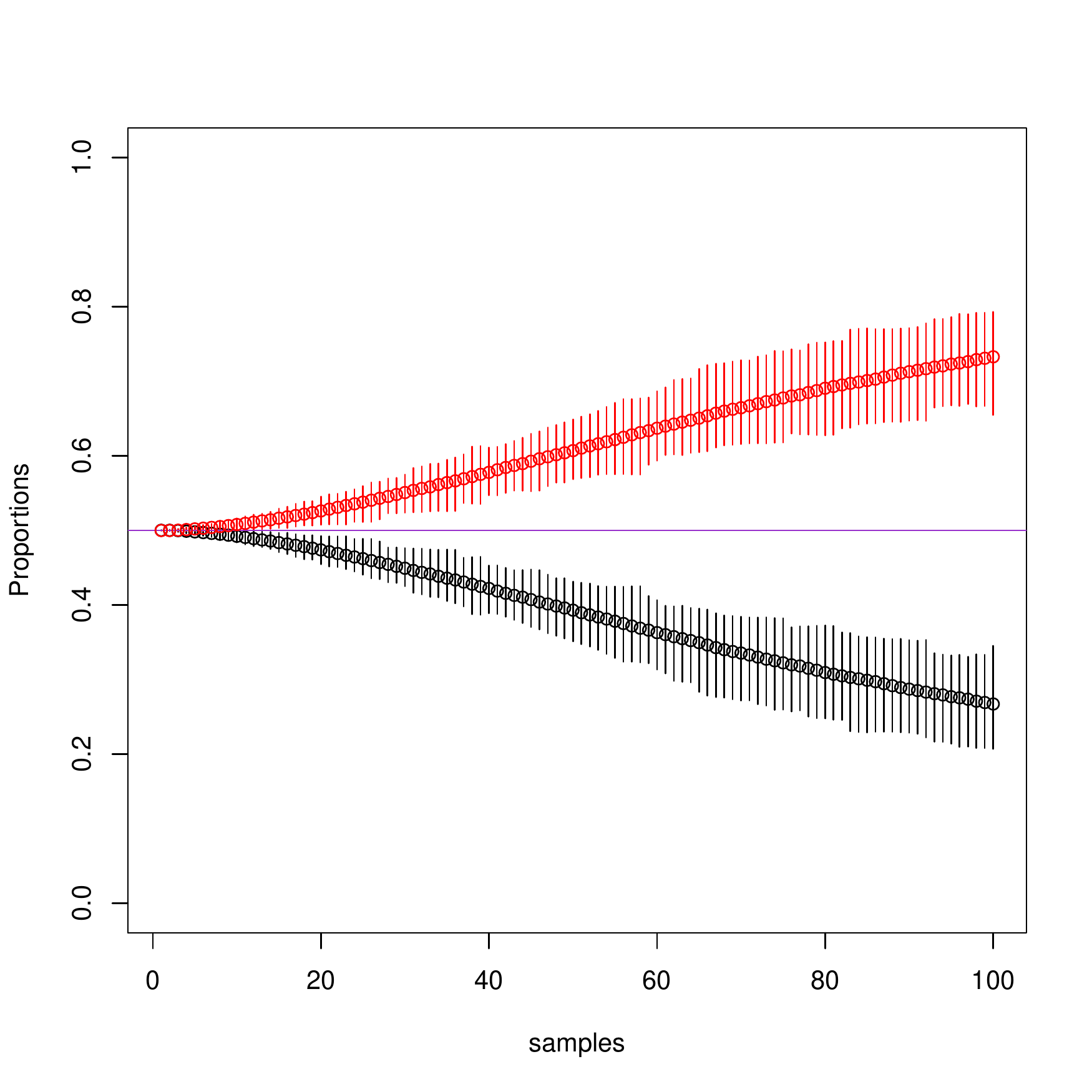}}}\\
	\subfloat[$\delta_t=5, \omega_t = 0.1$.]{%
		\resizebox*{4cm}{!}{\includegraphics{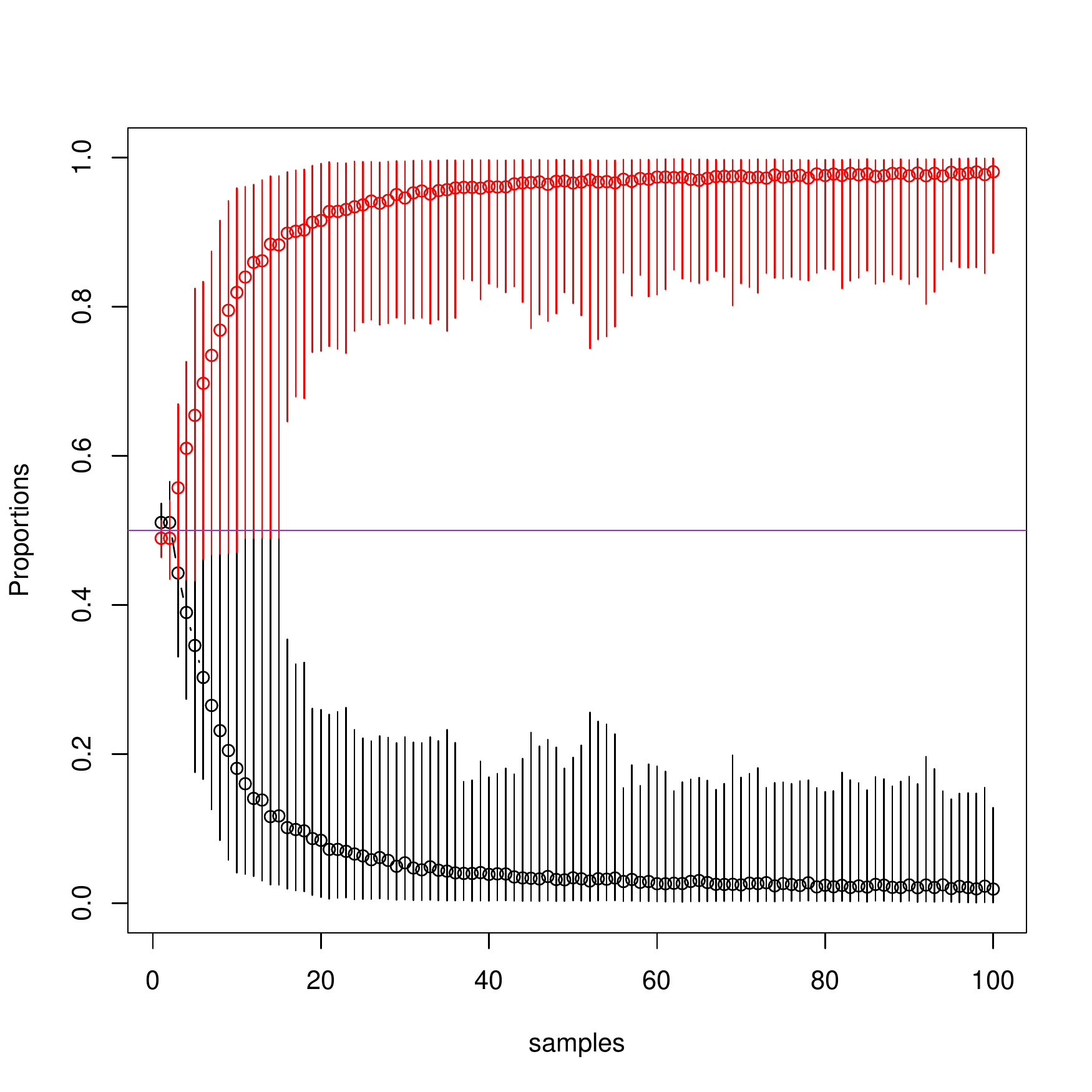}}}\hspace{5pt}
	\subfloat[$\delta_t=5, \omega_t = 0.01$.]{%
		\resizebox*{4cm}{!}{\includegraphics{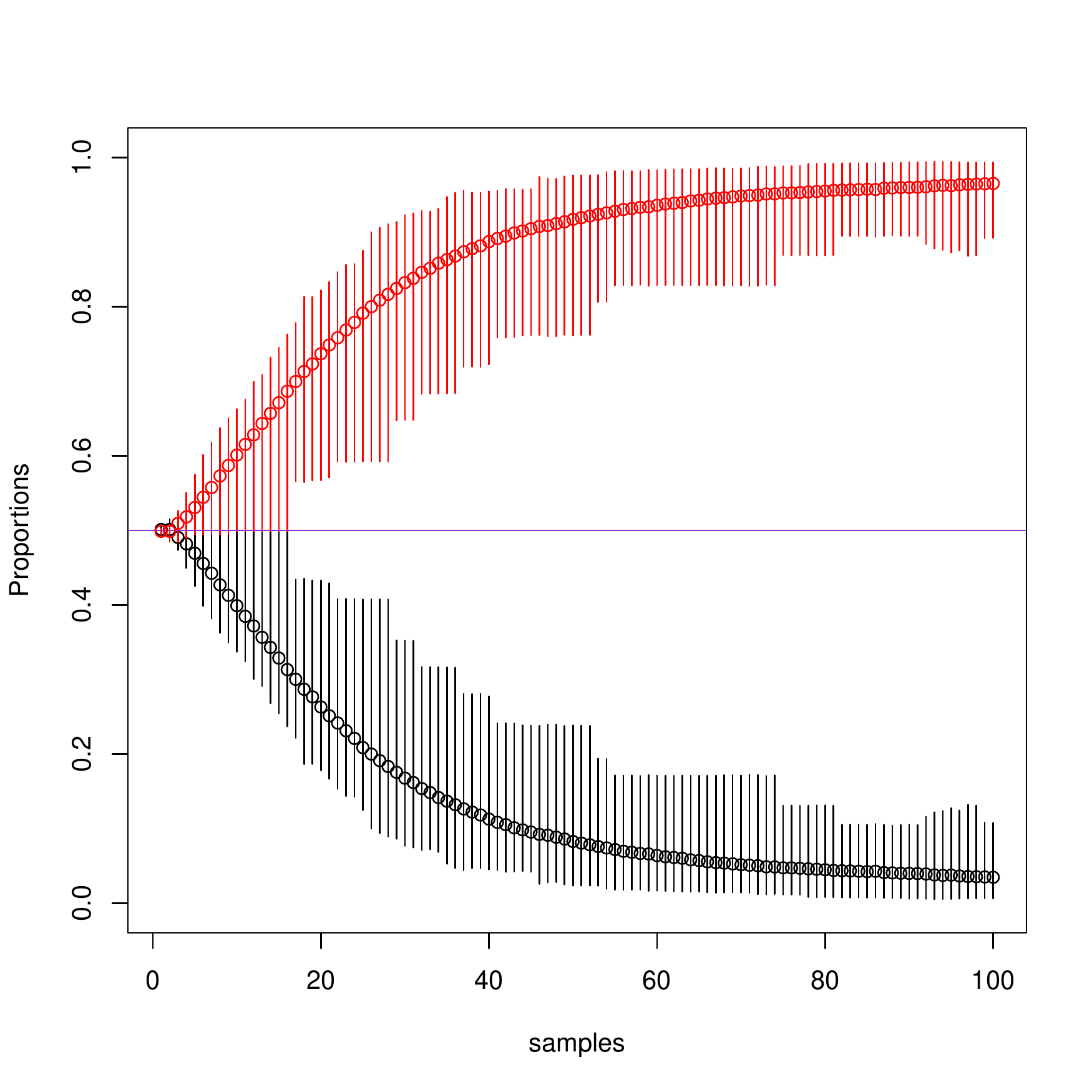}}}\hspace{5pt}
	\subfloat[$\delta_t=5, \omega_t = 0.001$.]{%
		\resizebox*{4cm}{!}{\includegraphics{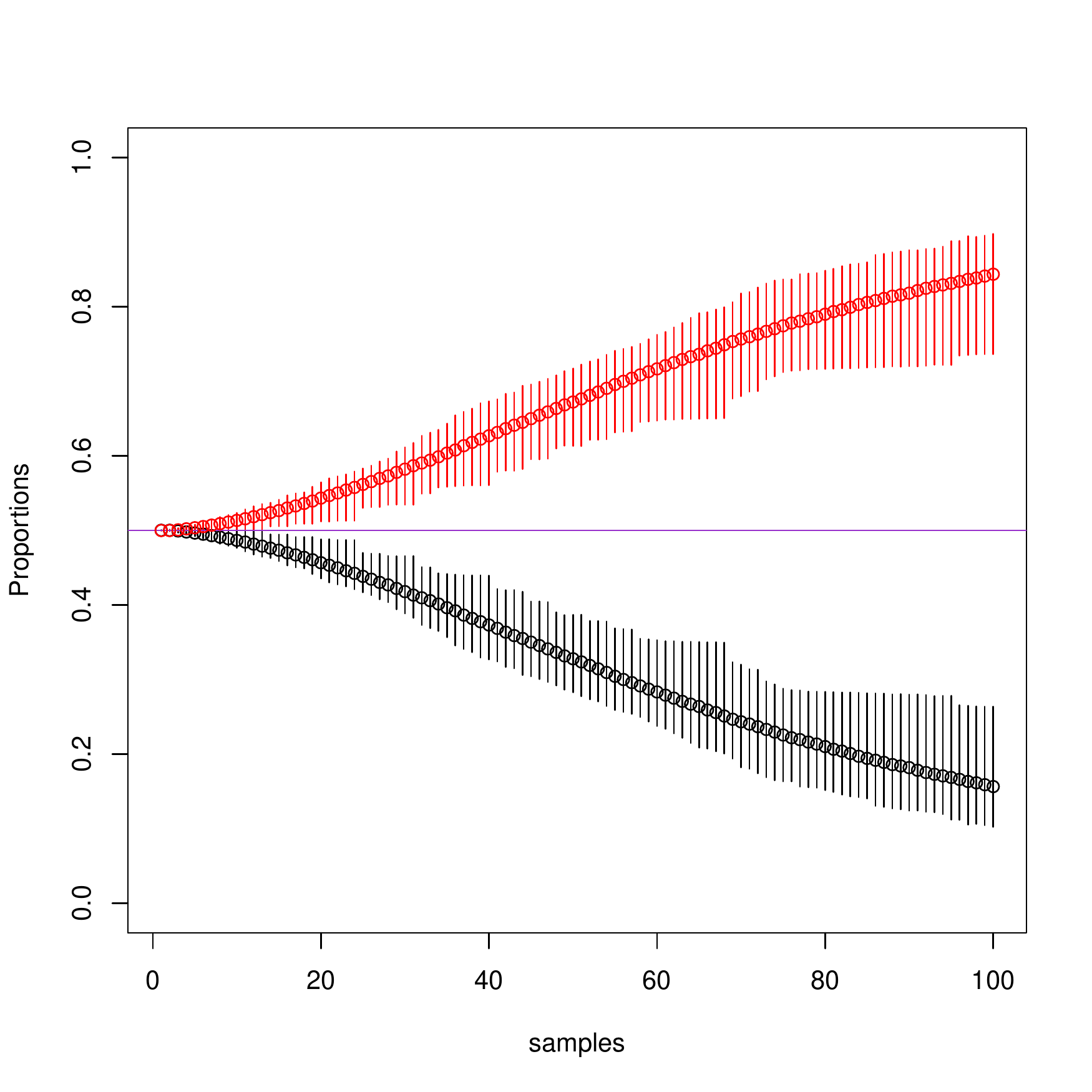}}}
	\caption{Comparison of Weight Allocation proportions for $\omega_{t} = 0.1, 0.01$ and $0.001$ and $\delta_t = 1, 3, 5$, $C_{t_B} =0.000001$ and $\beta_{t} = 1$ with bars representing the uncertainty across simulations.} \label{fig:compweightcov}
\end{figure}

A budget size of $\mathscr{N} = 100$ was chosen  and a sensitivity analysis was conducted using various values for $\delta_{t}$, $\omega_t$, and $c_{t_B}$, while keeping $Q_t = 1$. The values chosen for $\mu_B$ were 1 - 5, leading to $H_{A}: \delta_{t} = 1$ through $H_{A}: \delta_{t} = 5$. This lead to the hypothesis
\begin{center}
	\begin{eqnarray}  \label{eq:7}
	H_{0}: \delta_{t} &=& 0 \cr
	H_{A}: \delta_{t} &\ne& 0 \label{eq:7}
	\end{eqnarray} \label{eq:36}
\end{center}
where $\delta_{t} = \mu_{A} - \mu_{B}$ such that $\delta_{t} = {1, 2, 3, 4, 5}$. Furthermore, $\omega_t = 0.1, 0.01, 0.001$, and $c_{t_B} = 0.1, 0.001, 0.000001$.   Decreasing the values for $\omega_{t}$ represents an increased certainty of between time variability impact. Finally, decreasing the values of $c_{t_B}$ results in an increased knowledge group B has no effect. The weighted allocation proportion values in Figure~\ref{fig:compweightcov} represent each of the $\delta_t$ and $\omega_t$ values. However only the $c_{t_B} = 0.000001$ was chosen because it provides the best illustration of the impact seen in the sensitivity analysis. Additionally, by using $\mathscr{N} = 100$ and retaining $Q_t = 1$ throughout the sensitivity analysis the varied values of $\delta_{t}$ represent 1\% to a 5\% difference in the two treatments.


The first analysis used $\delta_t = 1$ with $\omega_t = 0.1$ and the results are shown in Figure~\ref{fig:compweightcov}a. Using these values treatment A had a  mean proportion of allocation of 0.603 with treatment B allocation proportion equal to 0.397.  The treatment allocation switch from B to A had a mean value of 39.595. When $\omega_t = 0.01$ in Figure~\ref{fig:compweightcov}b the mean proportion of allocation values to treatment A decreased to 0.595, while treatment B allocation increased to 0.405. However, the mean allocation switch from B to A increased slightly from 39.595 to 40.913. Finally, Figure~\ref{fig:compweightcov}c provides the results when letting $\omega_t = 0.001$. Here the mean proportion of allocation values to treatment A was 0.538 with that allocated to treatment B equal to 0.462. Using $\delta_t = 1$ and patient entry time variances this accurate lead to the highest mean treatment allocation switch from B to A,  46.702.

Next $\delta_t = 3$ was chosen and the analysis was conducted. Using $\omega_t = 0.1$ treatment A had a  mean proportion of allocation of 0.793, with treatment B allocation proportion equal to 0.207, seen in Figure~\ref{fig:compweightcov}d. The treatment allocation switch from B to A had a mean value which decreased from 39.595 using $\delta_t = 1$ to 18.217 using $\delta_t = 3$. When $\omega_t = 0.01$, seen in Figure~\ref{fig:compweightcov}e, the mean proportion of allocation values to treatment A decreased to 0.751, while treatment B allocation increased to 0.249. However, the mean allocation switch from B to A decreased from  40.913 at $\delta_t = 1$ to 24.694 using $\delta_t = 3$. Finally, Figure~\ref{fig:compweightcov}f shows the results when $\omega_t = 0.001$.  Here the mean proportion of allocation values to treatment A decreased to 0.609 with mean proportion allocated to treatment B equal to 0.391. Once again the mean number at which the treatment allocation switched from B to A decreased from 46.702 using $\delta_t = 1$ to 39.499 using $\delta_t = 3$. 

Finally $\delta_t = 5$ was analyzed using the varied $\omega_{t}$ values. Using $\omega_t = 0.1$ treatment A had a  mean proportion of allocation of 0.885, with treatment B allocation proportion equal to 0.115, seen in Figure~\ref{fig:compweightcov}g. The mean number at which treatment allocation went from B to A was 8.159, which is much lower that the mean values for $\omega_t = 0.1$ when using $\delta_t = $ 1 or 3. When $\omega_t$ was decreased to 0.01, seen in in Figure~\ref{fig:compweightcov}h, the mean proportion of allocation values to treatment A decreased slightly to 0.833, while treatment B increased to 0.167. The mean number at which treatment allocation switched from A to B was 15.453, almost double that obtained using $\omega_{t} = 0.1$. Lastly, Figure~\ref{fig:compweightcov}i. shows the allocation weights when the value for  $\omega_t$ was chosen to be 0.001. Here treatment A had a mean allocation proportion allocation of the mean proportion of 0.665, while treatment B had a mean allocation proportion of 0.335, seen in Figure~\ref{fig:compweightcov}o. Using the more precise patient entry time variances, treatment allocation switched from B to A was 33.482, which is double the value at $\omega = 0.01$ and 4 times that when $\omega = 0.1$. 

By decreasing the value of $\omega_{t}$ within each $\delta_{t_i}$, it can be seen the mean allocation probabilities to treatment A decrease within each group, leading to lower convergent values in each $\delta_t$ group, i.e. when $\delta_{t} = 1$ the mean convergent values for treatment A go from 0.603, 0.595, 0.538 as information regarding $\omega_{t}$ became more precise. This indicates a higher treatment B allocation proportion. However, increasing $\delta_t$ also leads to an increased mean number of necessary allocations for more precise $\omega_{t}$. For instance, when $\delta_t = 3$, the number of allocation values are 18.217, 24.694, and 39.499 for $\omega = 0.1, 0.01, 0.001$ respectively. However, when each of the allocation values are compared with comparable values of  $\omega_{t}$ at each $\delta_t$ value, one may observe a diminished mean number for comparable values of $\omega_{t}$. For example, when allowing $\omega_{t} = 0.1$, the mean number goes from 39.595 at $\delta_t = 1$ to 18.217 at $\delta_t = 3$ to 8.159 when $\delta_t = 5$. In fact, it appears using $\delta_t = 5$ provides the lowest mean switching value at every comparable value of $\omega_{t}$.

\section{Stopping Rule}

In order to maintain a fully Bayesian approach to this research, a Bayes Factor was used to determine definitive results. Additionally the 95\% credible intervals and the associated medians were calculated and were used, along with the Bayes Factor, to determine when the algorithm flipped treatment assignment. 
In order to determine a \lq\lq{critical\rq\rq}  Bayes Factor value, \cite{kass1995bayes}, suggest using a Bayes Factor greater than 100 provides \lq\lq{Decisive evidence\rq\rq} against the null hypothesis of no difference.
However, the notation of \cite{gonen2005bayesian} was used for the calculation of the Bayes Factor, whereby the null hypothesis is in the numerator yielding
\begin{equation}  
p(H_{0} \mid \textbf(D)) = \frac{P(\textbf{D} \mid H_{0}) P(H_{0}))} {P(\textbf{D} \mid H_{0})p(H_{0})+P(\textbf{D} \mid H_{1})p(H_{1})}  
\end{equation} \label{eq:8}
In their definition, they have the null hypothesis in the numerator and this 
leads to the Bayes Factor
\begin{equation}  
BF_{01}= \frac{P(\textbf{D} \mid H_{0})}{P(\textbf{D} \mid H_{1})}
\end{equation} \label{eq:9bc}
Using equation 9, a Bayes Factor less than $\frac{1}{100}$ was chosen to provide \lq\lq{Decisive evidence\rq\rq}  and support towards the more favorable treatment. 

The Bayes Factor was calculated using the Bayesian Two Sample T-Test discussed in \cite{gonen2005bayesian}. They define the Bayes Two Sample T Test as
\begin{equation}  \label{eq:12}
BF = \frac{T_{\nu}(t \mid 0,1)}{T_{\nu}(t\mid n_{\delta}^{\frac{1}{2}}\lambda,1+n_{\delta}\sigma_{\delta}^2)}
\end{equation}
By choosing a Bayes Factor less than $\frac{1}{100}$ any Bayes Factor considered \lq\lq{Decisive\rq\rq} represented a 100 times more likely chance the allocation had switched. Any indecisive Bayes Factor indicated the budget size $\mathscr{N} = 100$ was exhausted and no treatment allocation switch had occurred. 
Parenthetical values in Table~\ref{tbl:budget123cov} and Table~\ref{tbl:budget45cov} represent median and 95\% credible interval values of the Bayes Factor while the bold numbers represent the Bayes Factor calculated at $\mathscr{N} = 100$.

Using the value $c_{t_B} = 0.000001$, a sensitivity analysis was conducted using $\delta_{t} = 1, 3$ and $5$ when varying $\omega_{t}$. These median, 95\% credible intervals, and Bayes Factors  may be seen in Table~\ref{tbl:budget123cov} and Table~\ref{tbl:budget45cov}. Any italicized Bayes Factor is considered highly decisive, and represents 100 times more likely a switch occurred. 

\begin{table}[ht]
	\caption{Covariate Included Budget Allocation N using $\delta_{t} = 1, 2, 3$ ($Q_{0.025}$, $Q_{0.5}$, $Q_{0.975}$) $\mathbf{P(N \ge 100)}$}
	\begin{center}
		\begin{adjustbox}{max width=\textwidth}
			\begin{tabular}{cc|rrr}\\
				\hline
				$$&$$&$$&$\delta_{t}$&$$\\
				\hline
				$C_t$&$\omega_t$&1&2&3\\
				\hline
				\multirow{3}{*}{0.1}&0.1&(25, 48, 84), \textit{\textbf{0.002}}&(22, 32, 51), \textit{\textbf{0.000}}&(22, 28, 45.025), \textit{\textbf{0.001}}\\
				&0.01&(47, 72, 100), \textbf{0.106}&(44, 55, 74), \textit{\textbf{0.000}}&(48, 56, 69.025), \textit{\textbf{0.000}}\\
				&0.001&(99, 100, 100),\textbf{0.974}&(100,100,100),\textbf{0.985}&(76, 90, 100),\textbf{0.166}\\
				\hline
				\multirow{3}{*}{0.001}&0.1&(27, 49, 86.025), \textit{\textbf{0.009}}&(21, 31, 48), \textit{\textbf{0.000}}&(23, 28, 46), \textit{\textbf{0.000}}\\
				&0.01&(50, 73, 100), \textbf{0.105}&(46, 58, 75), \textit{\textbf{0.000}}&(49, 57, 68.025), \textit{\textbf{0.000}}\\
				&0.001&(100,100,100),\textbf{1.000}&(100,100,100),\textbf{1.000}&(98,100,100),\textbf{0.956}\\
				\hline
				\multirow{3}{*}{0.000001}&0.1&(26.975, 47, 87), \textit{\textbf{0.009}}&(23, 32, 52), \textit{\textbf{0.000}}&(22, 27.5, 43), \textit{\textbf{0.000}}\\
				&0.01&(50, 74, 100), \textbf{0.114}&(47, 57, 76), \textit{\textbf{0.001}}&(49.975, 57, 69), \textit{\textbf{0.000}}\\
				&0.001&(100, 100, 100),\textbf{1.000}&(100, 100, 100),\textbf{1.000}&(99, 100, 100),\textbf{0.974}\\
				\hline 
			\end{tabular}
		\end{adjustbox}
	\end{center}
	\label{tbl:budget123cov}
\end{table}

Notice at $\omega_{t} = 0.1$ the Bayes factor for $\delta_{t} = 1$ is 0.009 (median =47, 95\% credible interval (26.975, 87)), while for $\delta_{t} = 2$ the Bayes Factor is 0.000 (median = 32, 95\% credible interval (23, 52)). When analyzing $\delta_{t} = 3$, a Bayes Factor of 0.000 was calculated (median  = 27.5, 95\% credible interval (22, 43)) An increase to $\delta_{t} = 4$ yielded a Bayes Factor of 0.010 (median = 28, 95\% credible interval (23, 73.025)). Each of these first 4 means indicated decisive evidence. However, when $\delta_{t} = 5$ the Bayes factor is 0.088 (median = 32, 95\% credible interval (25, 100)), indicating indecisive evidence suggesting no switch to the better treatment occurred prior to exhausting the patient budget size.
\begin{table}[ht]
	\caption{Budget Allocation using N using $\mu_{B} = 4, 5$ ($Q_{0.025}$, $Q_{0.5}$, $Q_{0.975}$) $\mathbf{P(N \ge 100)}$}
	\begin{center}
		\begin{tabular}{cc|rrr}\\
			\hline 
			$$&$$&$\mu_B$$$\\ 
			\hline 
			$C_t$&$\omega_t$&4&5\\
			
			\hline
			\multirow{3}{*}{0.1}&0.1&(22.000, 29.000, 78.025),\textit{\textbf{0.010}}&(25.000, 32.000,100.000), \textbf{0.113}\\
			&0.01&(50.000, 61.000, 76.000), \textit{\textbf{0.001}}&(45.000, 56.000, 85.000), \textit{\textbf{0.002}}\\
			&0.001&(63.000, 75.000, 94.000), \textbf{0.011}&(57.000, 68.000, 87.000), \textit{\textbf{0.000}}\\
			\hline
			\multirow{3}{*}{0.001}&0.001&(24.000, 28.000, 68.025), \textit{\textbf{0.006}}&(26.000, 31.000, 100.000), \textbf{0.074}\\
			&0.01&(50.000, 60.000, 73.000), \textit{\textbf{0.000}}&(46.000, 55.000, 73.000), \textit{\textbf{0.000}}\\
			&0.001&(86.000, 94.000, 100.000), \textbf{0.203}&(78.000, 86.000, 99.000), \textbf{0.021}\\
			\hline
			\multirow{3}{*}{0.000001}&0.1&(23.000, 28.000, 73.025), \textit{\textbf{0.010}}&(25.000, 32.000, 100.000), \textbf{0.088}\\
			&0.01&(49.000, 61.000, 76.000), \textit{\textbf{0.000}}&(45.000, 56.000, 76.025), \textit{\textbf{0.002}}\\
			&0.001&(85.000 ,94.000, 100.000), \textbf{0.223}&(79.000, 87.000, 100.000), \textbf{0.027}\\
			\hline
		\end{tabular}
	\end{center}
	\label{tbl:budget45cov}
\end{table}

When $\omega_{t}$ was reduced to 0.01, using $\delta_{t} = 1$ a Bayes Factor of 0.114 (median = 74, 95\% credible interval (50, 100)) was calculated indicating no decisive evidence of preferred treatment was found by $\mathscr{N} = 100$. Yet, when $\delta_{t} =2$ a Bayes Factor of 0.001 (median = 57, 95\% credible interval (47, 76)) indicated decisive evidence. Decisive evidence was also seen when $\delta_{t} =3$ with its Bayes Factor of 0.000 (median = 57, 95\% credible interval (49.975, 69)). Interestingly, using the value of $\delta_{t} = 4$ and $\delta_{t} = 5$  yielded Decisive Bayes Factors (0.000 and 0.002 respectively), which provided highly decisive evidence the allocation to the better treatment had occurred. However, using $\delta_{t} = 4$ median value was 61 (95\% credible interval 49, 76.000), while with $\mu_{B} = 5$ a lower median value of 56 was observed with 95\% credible interval (45, 76.025).

Lastly, when $\omega = 0.001$ was analyzed, the Bayes Factor for $\delta_{t} = 1$ and $\delta_{t} = 2$ were the same; a value of 1.000 which indicated no decisive evidence was found. Additionally, each had the same median and 95\% credible interval values of 100. The Bayes Factor decreased to 0.974 (median = 100, 95\% credible interval (99, 100)) when $\delta_{t} = 3$, however this was also indecisive. Likewise, even though the Bayes Factors decreased dramatically when $\delta_{t} = 4, 5$ (Bayes Factors of  0.223 and  0.027 respectively), no decisive evidence was found with these means either. However, when $\delta_{t} = 4$ median value was 94 with 95\% credible interval values (85, 100), however, when $\mu_{B} = 5$, median value was 87, with 95\% credible interval values (79, 100).

This analysis provides insight into how researchers may plan patient budget sizes. When choosing mean difference values between 1 and 3, it appears as though using the mean difference of 3 provides the lowest median and credible interval values using low to moderate belief in the variability between patients. However, it appears as though $\delta_{t} = 5$ provides the lowest median value when using a moderate variance of $\omega_{t} = 0.01$ using  $c_{t_B} = 0.000001$.  Furthermore, when using the highest variance accuracy of $\omega_{t} = 0.001$, there is no Decisive evidence shown for any choice of mean at $c_{t_B} = 0.000001$. Likewise, using $\omega_{t} = 0.01$ showed Decisive evidence for all mean values except $\delta_{t} = 1$

\begin{figure}[ht]
	\centering
	\subfloat[$\mu_{B}=1, \beta_t = 1$.]{%
		\resizebox*{4cm}{!}{\includegraphics{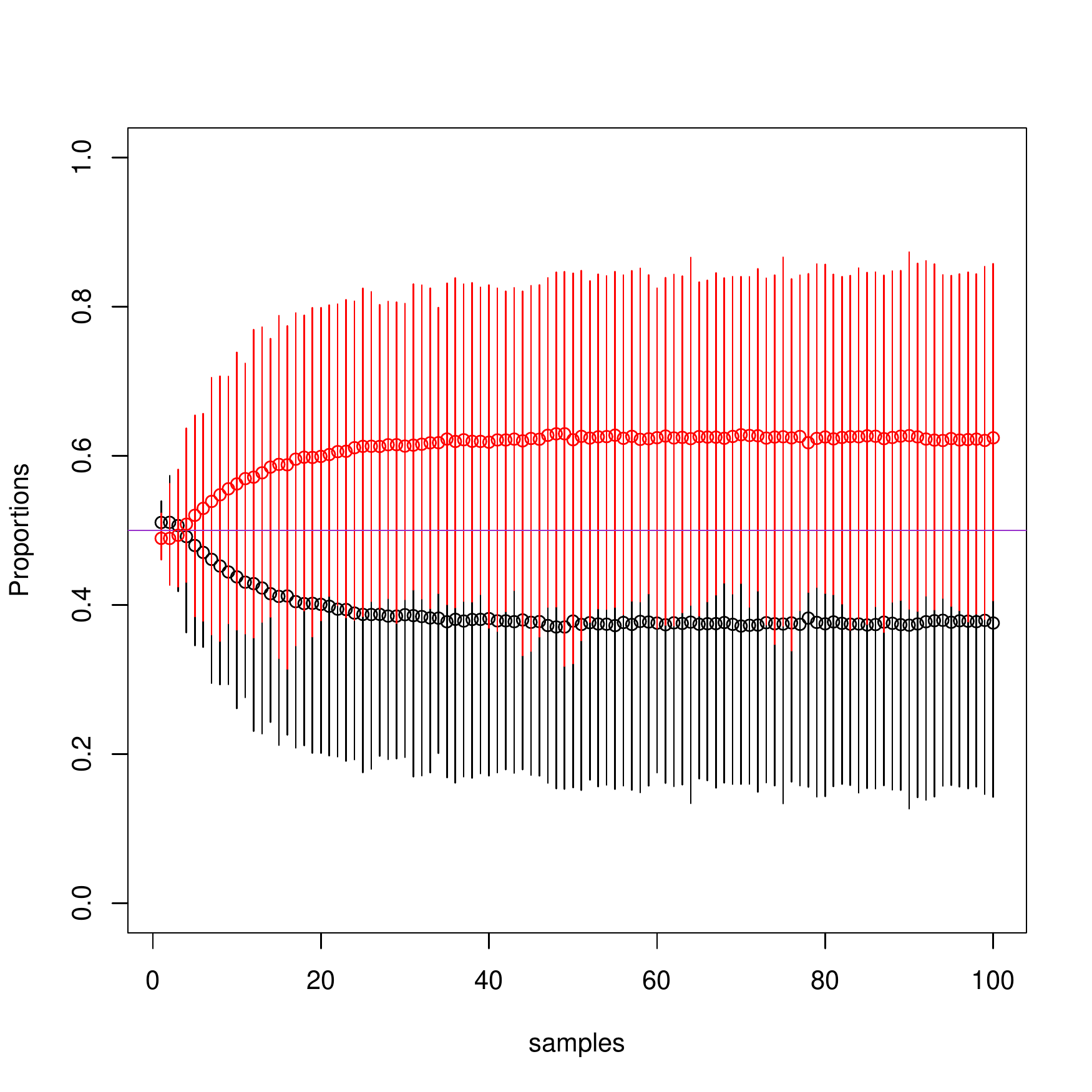}}}\hspace{5pt}
	\subfloat[$\mu_{B}=1, \beta_t = 2$.]{%
		\resizebox*{4cm}{!}{\includegraphics{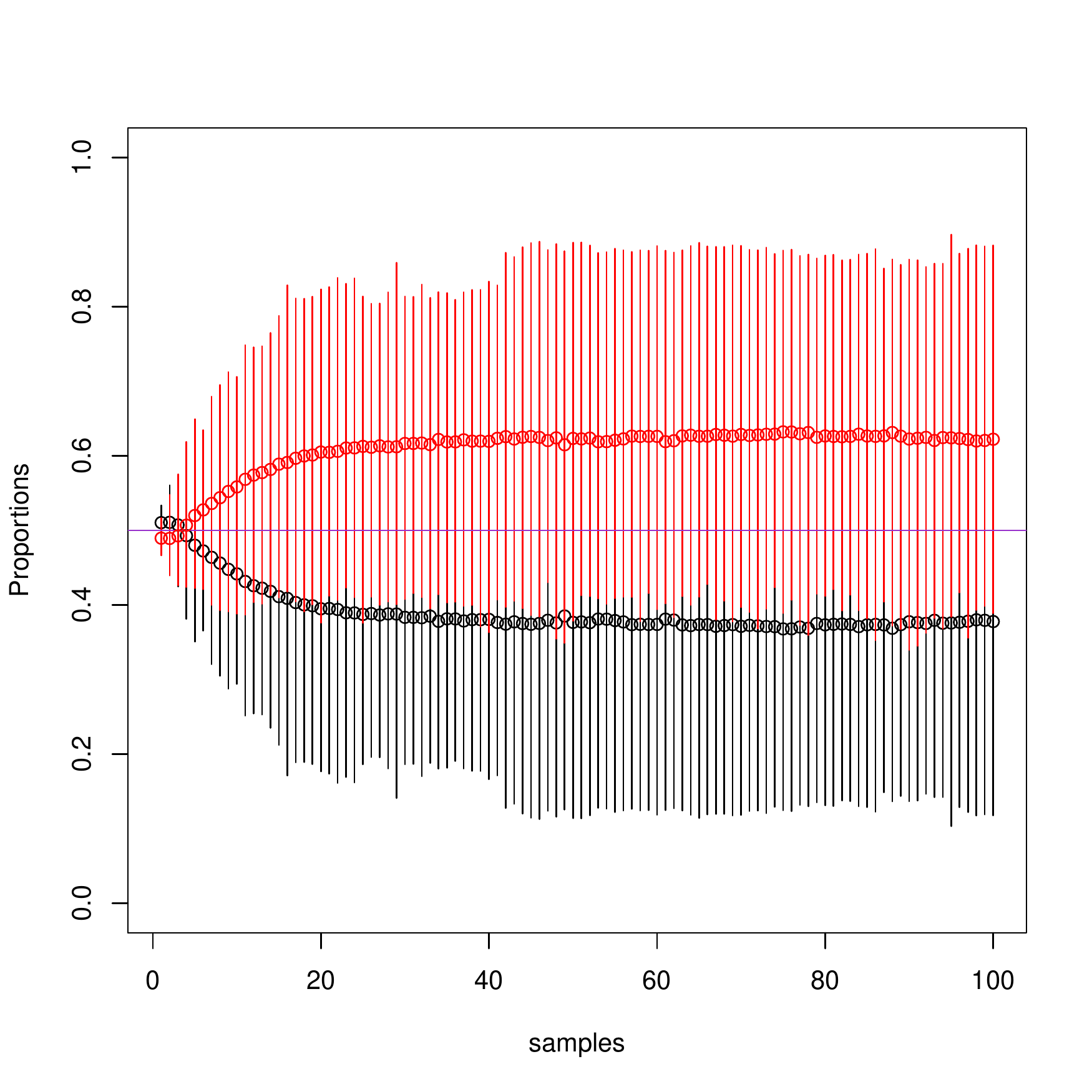}}}\hspace{5pt}\\
	\subfloat[$\mu_{B}=3, \beta_t = 1$.]{%
		\resizebox*{4cm}{!}{\includegraphics{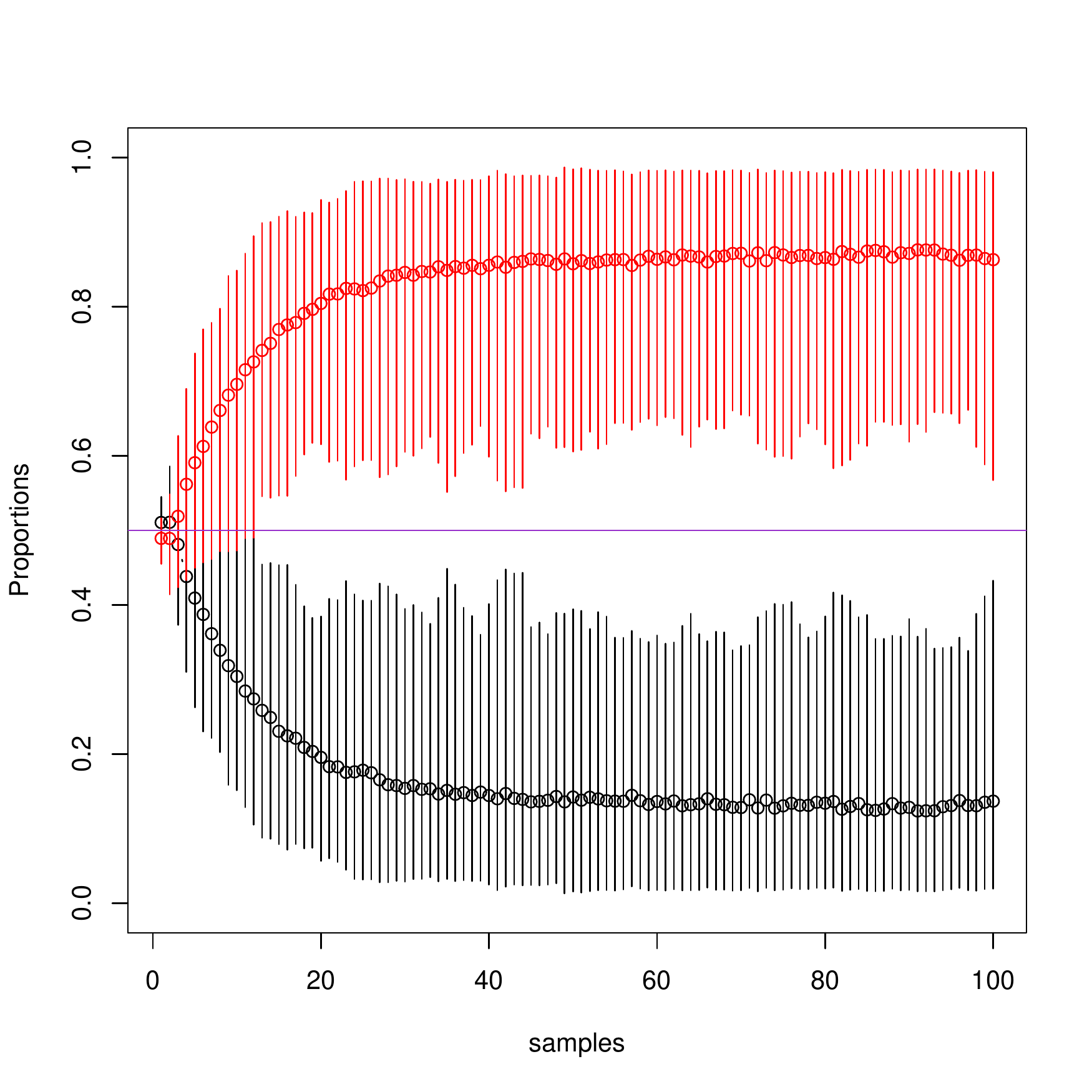}}}\hspace{5pt}
	\subfloat[$\mu_{B}=3, \beta_t = 2$.]{%
		\resizebox*{4cm}{!}{\includegraphics{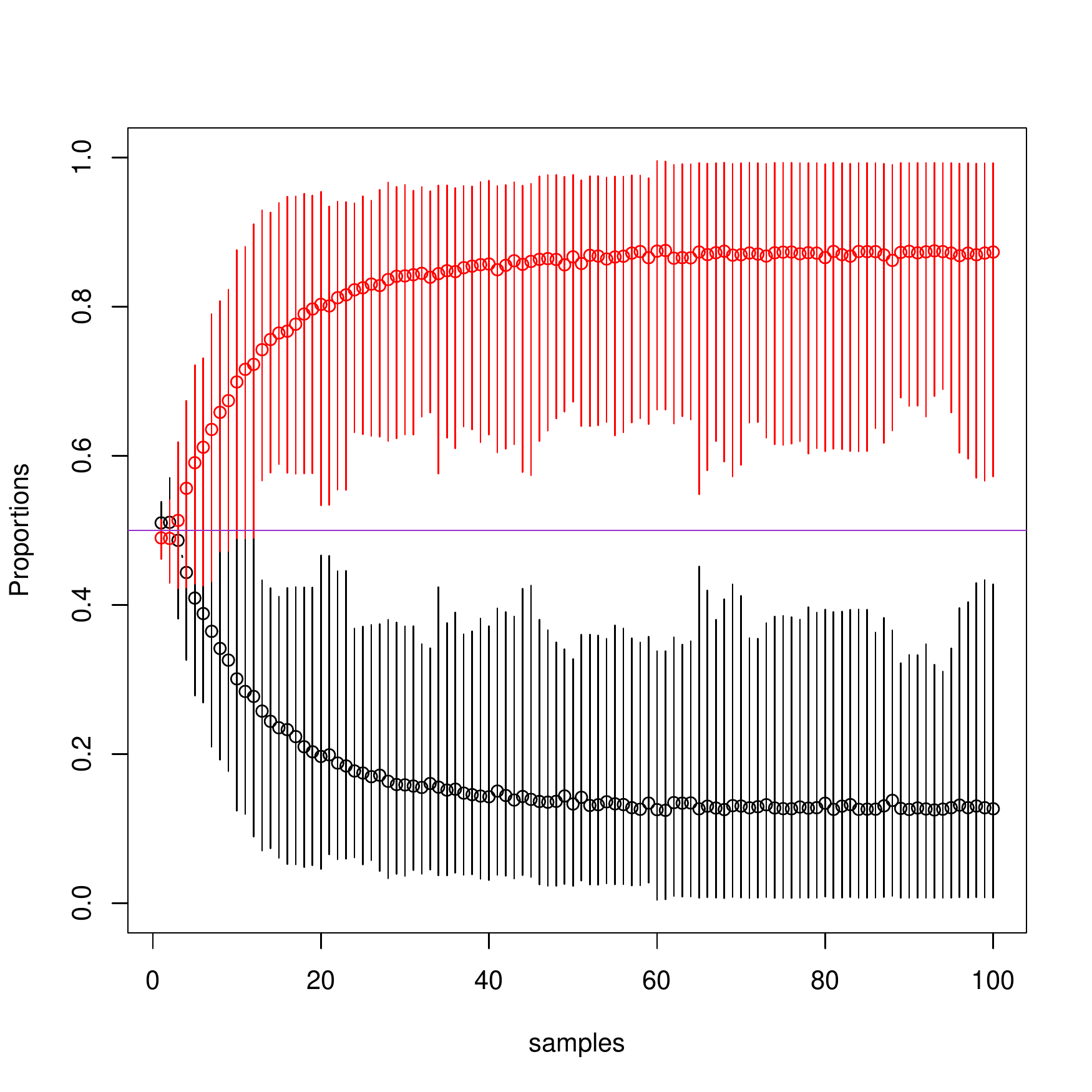}}}\hspace{5pt}\\
	\subfloat[$\mu_{B}=5, \beta_t = 1$.]{%
		\resizebox*{4cm}{!}{\includegraphics{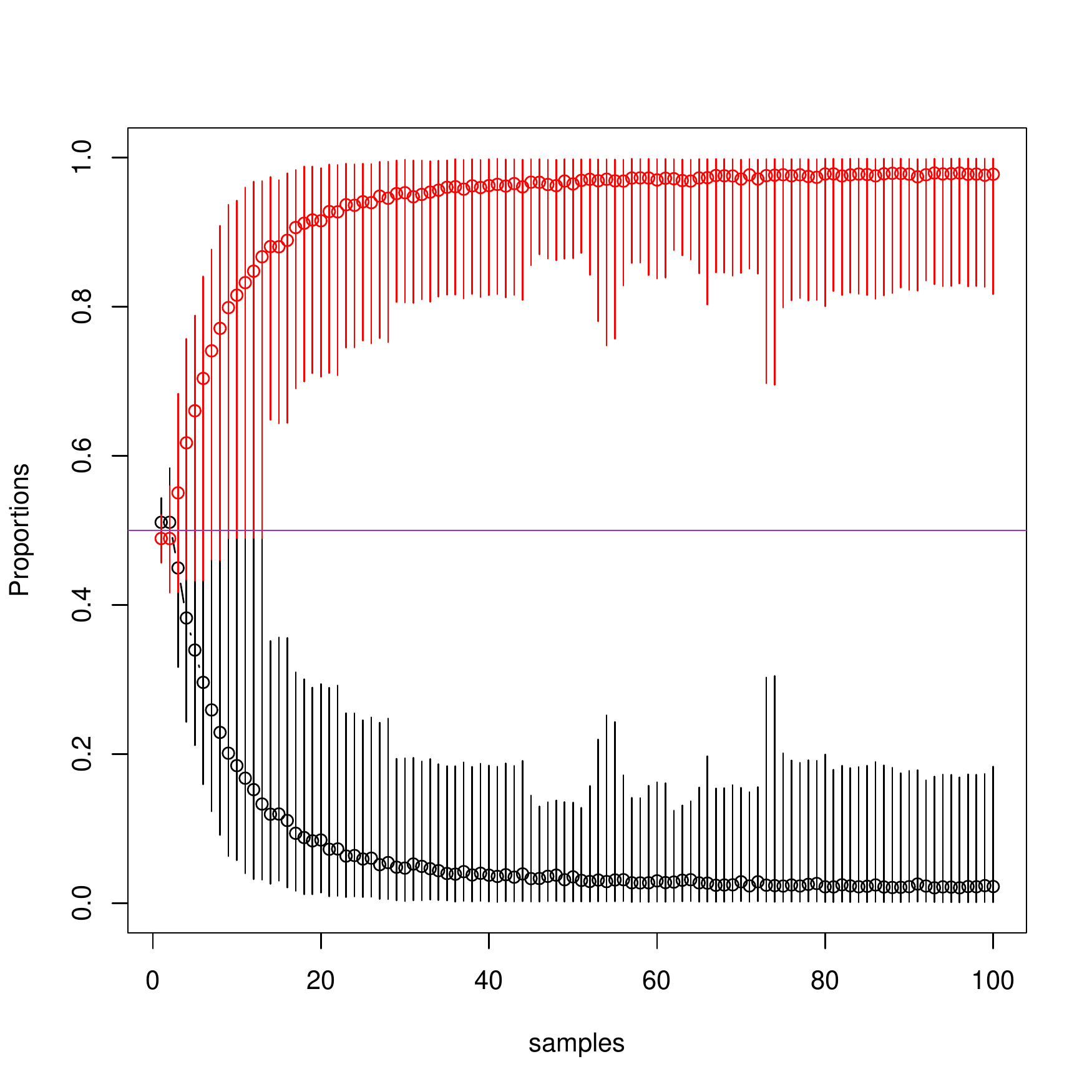}}}\hspace{5pt}
	\subfloat[$\mu_{B}=5, \beta_t = 2$.]{%
		\resizebox*{4cm}{!}{\includegraphics{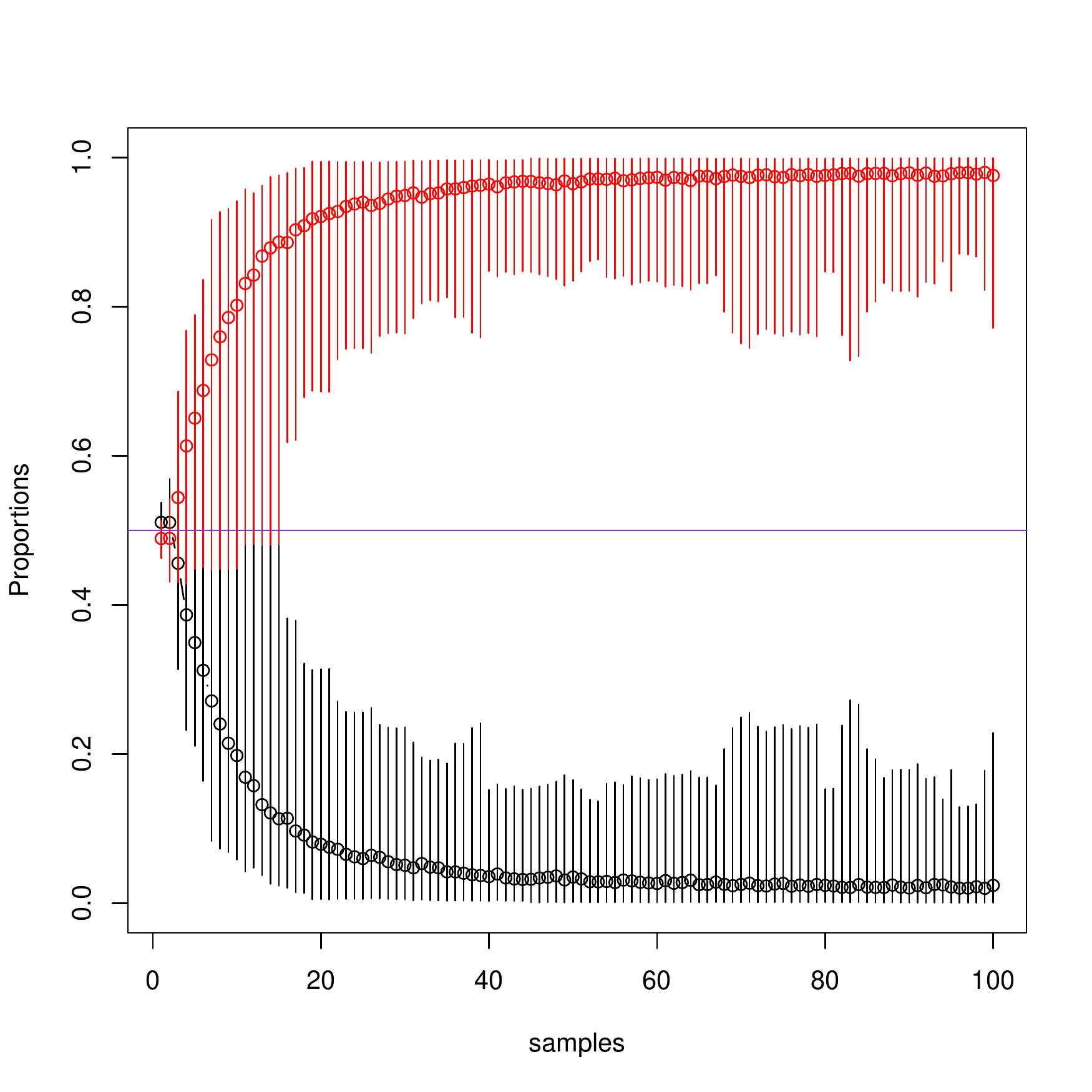}}}\hspace{5pt}
	\caption{Weight Allocation Proportion Comparisons when  $\mu_{B} = 1, 3, 5$, $\omega_{t} = 0.1$, $c_{t_B} = 0.000001$ and $\beta = 1, 2$ The bars represent the uncertainty across simulations} \label{fig:betacovcom}
\end{figure}


\section{Covariate  Comparison}

A comparison of the mean proportional allocation to treatments was also conducted when using values of $\beta = 1, 2$. This comparison was conducted using $\delta_{t} = 1,3,5$, $\omega_{t} = 0.1$ and $c_{t_B} = 0.000001$. The results when comparing $\beta = 1$ and $\beta = 2$ for each of the $\delta_{t},\omega_{t},c_{t_B}$ may be seen in Figure~\ref{fig:betacovcom} and the results of the Bayes Factor may be seen in Table~\ref{tbl:budget45covbeta12}.  For the graphs of $\delta_{t} = 1,2,3,4,5$, $\omega_{t} = 0.1$ and $c_{t_B} = 0.000001$, please see Figure~\ref{fig:betacovcomall} in Appendix A.

\begin{table}[ht]
	\caption{Budget Allocation using N using $\mu_{B} = 4, 5$ ($Q_{0.025}$, $Q_{0.5}$, $Q_{0.975}$) $\mathbf{P(N \ge 100)}$ for $\beta_{t} = 1, 2$}
	\begin{center}
		\begin{tabular}{c|rrr}\\
			\hline 
			$$&$\beta_{t}$&$$\\ 
			\hline 
			$\delta_t$&1&2\\
			
			\hline
			1&(27.000, 50.000, 86.025), \textit{\textbf{0.008}}&(26.000, 49.000, 88.025), \textit{\textbf{0.007}}\\
			\hline
			3&(23.000, 28.000, 43.025), \textit{\textbf{0.000}}&(21.000,    28.000, 44 ), \textit{\textbf{0.000}}\\
			\hline
			5&(26.000,  32.000, 100.000), \textbf{0.083}&(26.000,  31.000, 100.000), \textbf{0.084}\\
			\hline
		\end{tabular}
	\end{center}
	\label{tbl:budget45covbeta12}
\end{table}
When using $\delta_{t} = 1$ and $\beta_{t} = 1$ the mean proportion allocation for treatment A is 0.608, while the mean proportion allocation for treatment B is 0.392. However, when $\beta_{t}$ is increased to 2, the mean proportion allocation for Treatment A shows only a slight decrease to 0.604 while the mean proportion allocation for treatment B is 0.396, illustrating when using $\delta_{t} = 1$ there is very little change in allocation proportion based on the change in covariate value from 1 to 2. Likewise, the allocation switch from B to A using $\beta_{t} = 1$ was 39.813, with a similar value of 39.554 when using $\beta_{t} = 2$, indicating changing the covariate value had little impact on when the switch from B to A occurred. 

When using $\delta_{t} = 3$ and $\beta_{t} = 1$ the mean proportion allocation for treatment A is 0.791, while the mean proportion allocation for treatment B is 0.209. However, when $\beta_{t}$ is increased to 2, these allocation proportions remain the same illustrating when using $\delta_{t} = 3$ there is no change in allocation proportion when the covariate value increases from 1 to 2. Likewise, the allocation switch from B to A using $\beta_{t} = 1$ was 18.568, with a similar value of 18.544 when using $\beta_{t} = 2$. This again indicates increasing the covariate value from 1 to 2 has little to no impact on when the algorithm will switch from B to A.

When using $\delta_{t} = 5$ and $\beta_{t} = 1$ the mean proportion allocation for treatment A is 0.888, while the mean proportion allocation for treatment B is 0.112. However, when $\beta_{t}$ is increased to 2, the mean proportion allocation for Treatment A shows only a slight increase to 0.889 while the mean proportion allocation for treatment B is slightly decreased to 0.111, illustrating when using $\delta_{t} = 5$ there is very little change in allocation proportion when $\beta_{t}$ is increased from 1 to 2. Likewise, the allocation switch from B to A using $\beta_{t} = 1$ was 8.359, with a similar value of 8.375 when using $\beta_{t} = 2$. Similar to both $\delta = 1$ and $\delta = 3$ this indicates increasing the covariate value from 1 to 2 has little to no impact on when the algorithm will switch from B to A.

Finally, the median and 95\% credible intervals and Bayes Factor were calculated for for $\delta = 1, 3, 5$ and $\beta_{t} =1, 2$ and these may be seen in Table~\ref{tbl:budget45covbeta12}. Note that when $\delta =$ 1 and 3, decisive Bayes factors were found at both $\beta_{t} =$ 1 and 2. The median value for the combination $\delta_{t} =1$ and $\beta_{t} = 1$ was 50 (95\% credible interval 27.000, 86.025) with a Bayes factor of 0.008, while when $\beta_{t} = 2$ the median was 49.000 (95\% credible interval 26.000, 88.025) with a Bayes factor of 0.007. When $\delta_{t}$ is increased to 3, the median value decreases to 28 (95\% credible interval 23.000, 43.025) with a Bayes factor of 0.000 using $\beta_{t} = 1$, yet when $\beta_{t} = 2$, while the median value of 28 remains unchanged, the 95\% credible interval ranges from 21.000 to 44.000, and the Bayes factor value of 0.000 remains unchanged. Finally, when $\delta_{t}$ is increased to 5, a median value of 32.000 is found using $\beta_{t} = 1$ with 95\% credible interval 26.000, 100.00 and an indecisive Bayes factor of 0.083. When $\beta_{t}$ is increased to 2, the median value only slightly changes from 32 to 31, however, the 95\% credible interval values remain unchanged (26.000, 100.000), again with an indecisive Bayes factor of 0.084.
\subsection*{Conclusion}
Researchers conducting Bayesian Random Allocation models for clinical trials can be faced with computationally intensive problems when running large scale simulations requiring MCMC methods. These models are further complicated when a covariate is introduced. In the current application, a DLM was applied to random allocation models with a single covariate to demonstrate the ability to reduce time and patient allocation size in the presence of a covariate. Additionally, a sensitivity analysis was conducted both on mean proportion of allocation to each treatment and mean value required to switch to the preferred treatment. This provides insight for researchers who wish to know what treatment allocation proportion may be expected using varying difference values between $\mu_{A}$ and $\mu_{B}$, between time variances $\omega_{t}$ and current treatment B variance $c_{t_B}$, thereby providing insight into the different model behaviors.  Likewise, a power analysis was conducted using a Bayes Factor. This power analysis indicated the lowest median Bayes Factor occurred for a difference $\delta_{t} = 5$ using $\omega_{t} = 0.01$. This provides insight into necessary patient budget to determine a favorable treatment identification stopping criterion. This reduction of patient budget should reduce, if not eliminate the ethical issues caused by the increased unfavorable treatment allocation necessary using other allocation methods by allowing the more favorable treatment to be applied earlier in the clinical trial. Additionally, covariate values of 1 and 2 were analyzed using $\delta_{t} = 1,3,5$ while holding $\omega_{t} = 0.1$ and $c_{t_B} = 0.000001$ and a power analysis conducted using these values. It appears $\delta_{t} = 3$ provides the lowest median and the most decisive Bayes factor, regardless of which $\beta_{t}$ is chosen. Likewise, it appears that increasing $\beta_{t}$ from 1 to 2 has little to no impact on model performance. Future works may include a sensitivity analysis using multiple values for $\beta$ with larger differences than an increase of 1 unit. Additionally an examination of multi-arm studies with covariates, and survival analysis applications should be studied.

\section{References}
\pagebreak
\section{Appendix A.}
\begin{figure}[ht]
	\centering
	\begin{adjustbox}{max width=\columnwidth,totalheight=.75\textheight-2\baselineskip}
		\begin{tabular}{cc}
			\includegraphics[width=.33\textwidth]{allocwcov11000001beta1.pdf}& \includegraphics[width=.33\textwidth]{allocwcov11000001beta2.pdf}\\
			(a) $\mu_{B}=1, \beta_t = 1$ & (b) $\mu_{B}=1, \beta_t = 2$ \cr
			\includegraphics[width=.33\textwidth]{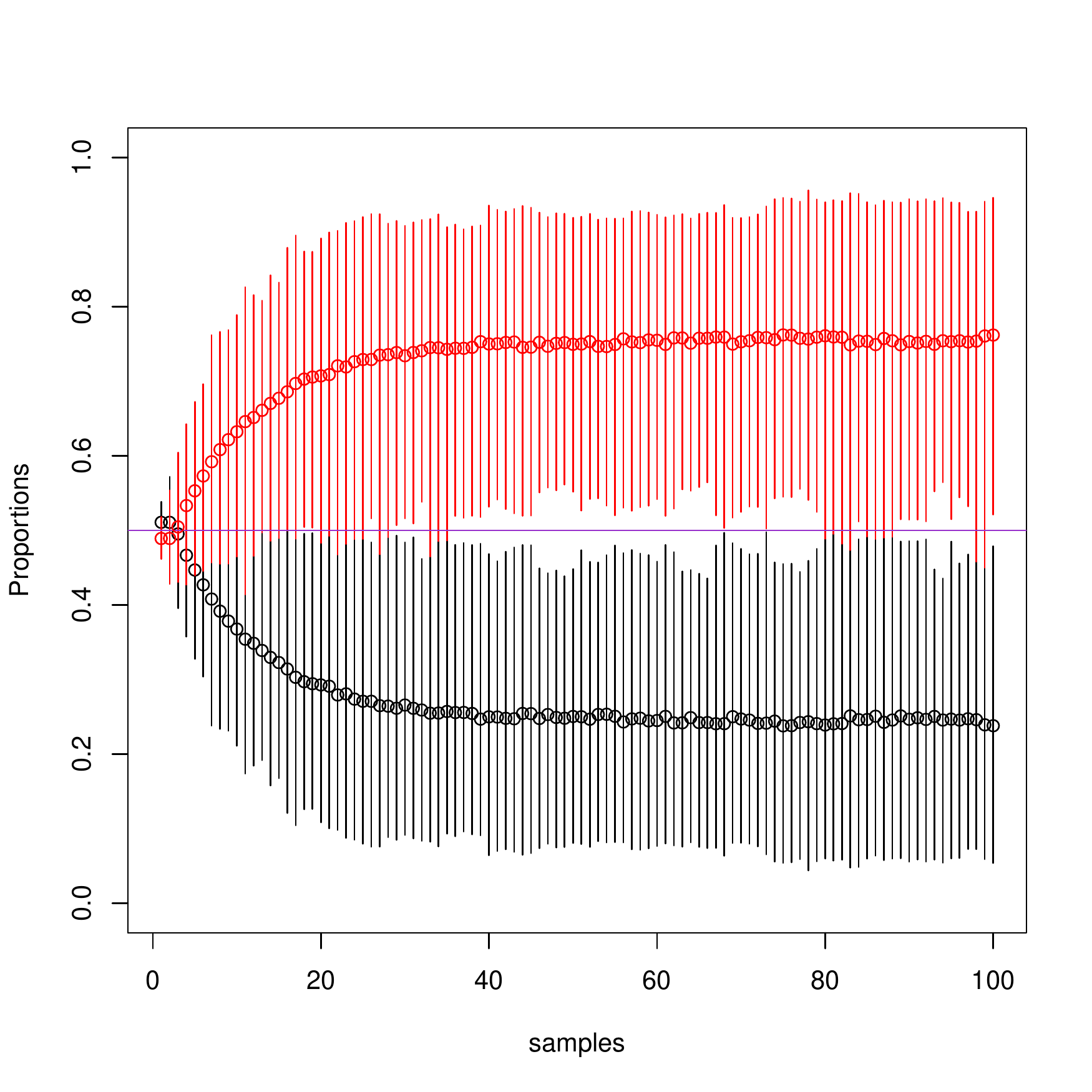}& \includegraphics[width=.33\textwidth]{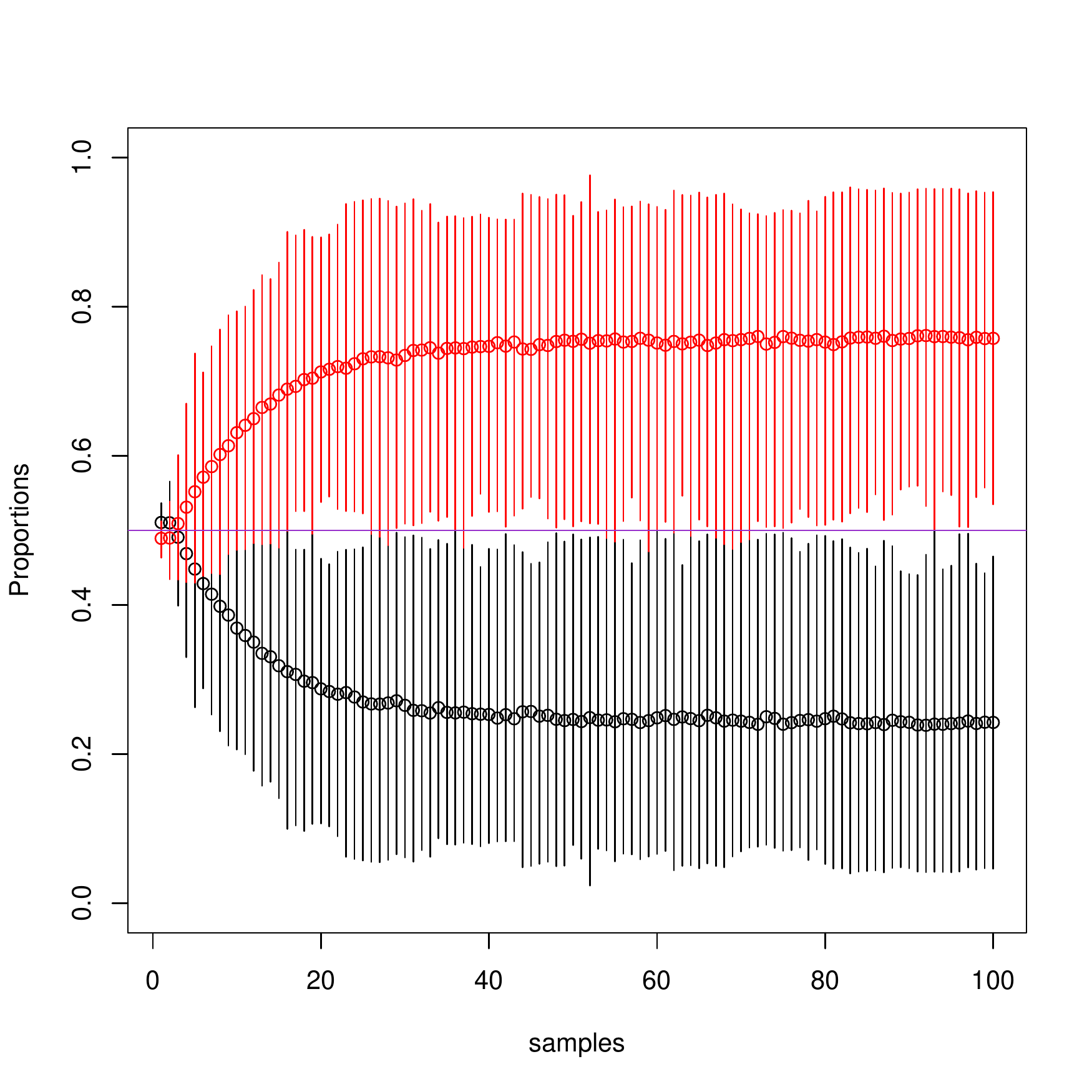}\\
			(c) $\mu_{B}=2, \beta_t = 1$ & (d) $\mu_{B}=2, \beta_t = 2$ \cr
			\includegraphics[width=.33\textwidth]{allocwcov31000001beta1.pdf}& \includegraphics[width=.33\textwidth]{allocwcov31000001beta2.pdf}\\
			(e) $\mu_{B}=3, \beta_t = 1$ & (f) $\mu_{B}=3, \beta_t = 2$ \cr
			\includegraphics[width=.33\textwidth]{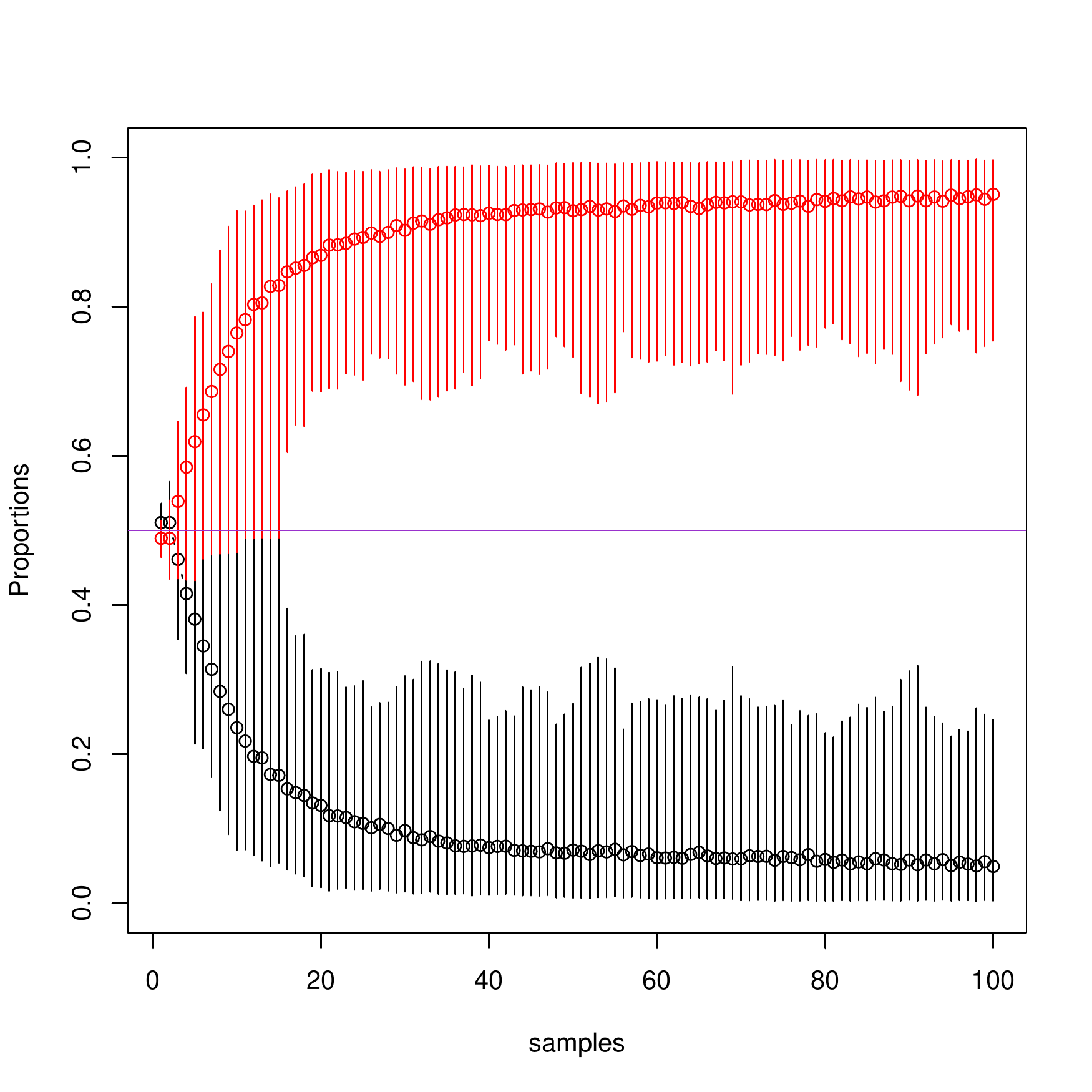}& \includegraphics[width=.33\textwidth]{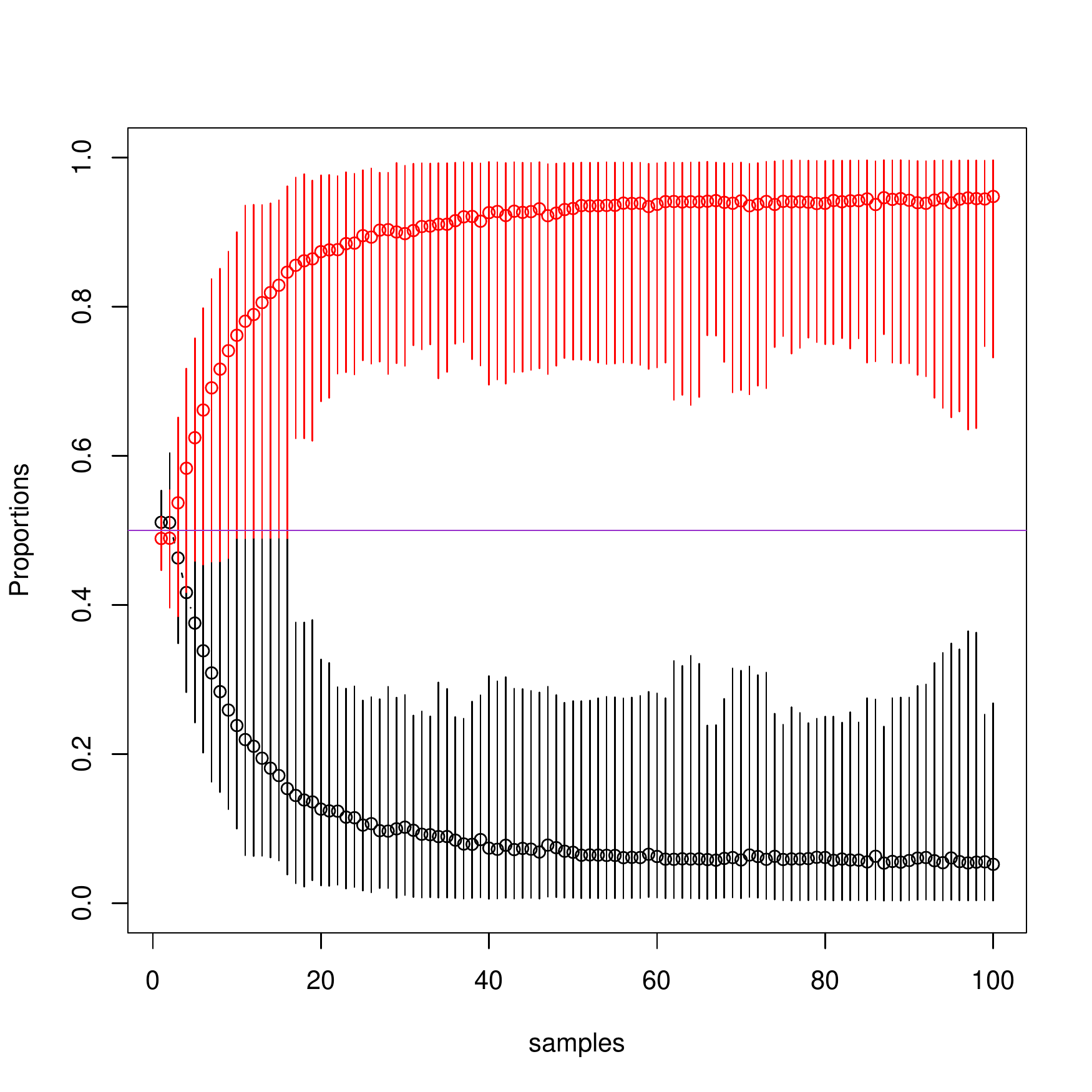}\\
			(g) $\mu_{B}=4, \beta_t = 1$ & (h) $\mu_{B}=4, \beta_t = 2$ \cr
			\includegraphics[width=.33\textwidth]{allocwcov51000001beta1.pdf}& \includegraphics[width=.33\textwidth]{allocwcov51000001beta2.pdf}\\
			(i) $\mu_{B}=5, \beta_t = 1$ & (j) $\mu_{B}=5, \beta_t = 2$\\
		\end{tabular}
	\end{adjustbox}
	\caption{Weight Allocation Proportion Comparisons when  $\mu_{B} = 1, 2, 3, 4, 5$, $\omega_{t} = 0.1$, $c_{t_B} = 0.000001$ and $\beta = 1, 2$ The bars represent the uncertainty across simulations}
	\label{fig:betacovcomall}
\end{figure}

\end{document}